\documentclass[aps,prb,superscriptaddress,nofootinbib,longbibliography,twocolumn]{revtex4-2}
\usepackage{bm}
\usepackage{lmodern}

\usepackage[utf8]{inputenc}
\usepackage{amsmath}
\usepackage{amssymb}
\usepackage{graphicx}
\usepackage{epstopdf}
\usepackage{xcolor}
\usepackage{enumerate}
\usepackage{ulem}
\usepackage[francais,english]{babel}
\usepackage{braket}
\usepackage{natbib}
\usepackage{float}
\usepackage{stmaryrd}

\usepackage[misc]{ifsym}

\newcommand\blankfootnote[1]{%
  \let\thefootnote\relax\footnotetext{#1}%
  \let\thefootnote\svthefootnote%
}

\usepackage{bibunits}

\usepackage{hyperref}
\usepackage{lipsum}

\begin{document}

\title{Probing resonating valence bonds on a programmable germanium quantum simulator}

\author{Chien-An Wang}
\altaffiliation{These authors contributed equally}
\affiliation{QuTech and Kavli Institute of Nanoscience, Delft University of Technology, PO Box 5046, 2600 GA Delft, The Netherlands}
\author{Corentin Déprez}
\altaffiliation{These authors contributed equally}
\author{Hanifa Tidjani}
\affiliation{QuTech and Kavli Institute of Nanoscience, Delft University of Technology, PO Box 5046, 2600 GA Delft, The Netherlands}
\author{William I. L. Lawrie}
\affiliation{QuTech and Kavli Institute of Nanoscience, Delft University of Technology, PO Box 5046, 2600 GA Delft, The Netherlands}
\author{Nico W. Hendrickx}
\affiliation{QuTech and Kavli Institute of Nanoscience, Delft University of Technology, PO Box 5046, 2600 GA Delft, The Netherlands}
\author{Amir Sammak}
\affiliation{QuTech and Netherlands Organisation for Applied Scientific Research (TNO), Delft, The Netherlands}
\author{Giordano Scappucci}
\affiliation{QuTech and Kavli Institute of Nanoscience, Delft University of Technology, PO Box 5046, 2600 GA Delft, The Netherlands}
\author{Menno Veldhorst}
\affiliation{QuTech and Kavli Institute of Nanoscience, Delft University of Technology, PO Box 5046, 2600 GA Delft, The Netherlands}
% \date{\today}

\begin{abstract}

\begin{center}
\vspace{-0.4cm}
% \Letter{~:~\href{mailto:C.C.Deprez@tudelft.nl}{C.C.Deprez@tudelft.nl},~\href{mailto:M.Veldhorst@tudelft.nl}{M.Veldhorst@tudelft.nl}}
\end{center}

\vspace{+0.3cm}
Simulations using highly tunable quantum systems may enable investigations of condensed matter systems beyond the capabilities of classical computers. Quantum dots and donors in semiconductor technology define a natural approach to implement quantum simulation. Several material platforms have been used to study interacting charge states, while gallium arsenide has also been used to investigate spin evolution. However, decoherence remains a key challenge in simulating coherent quantum dynamics. Here, we introduce quantum simulation using hole spins in germanium quantum dots. We demonstrate extensive and coherent control enabling the tuning of multi-spin states in isolated, paired, and fully coupled quantum dots. We then focus on the simulation of resonating valence bonds and measure the evolution between singlet product states which remains coherent over many periods. Finally, we realize four-spin states with $s$-wave and $d$-wave symmetry. These results provide means to perform non-trivial and coherent simulations of correlated electron systems. \linebreak

\end{abstract}

\maketitle

Quantum computers have the potential of simulating physics beyond the capacity of classical computers~\cite{Feynman1982,Lloyd1996,Abrams1997,Aspuru2005}. Gate-defined quantum dots are extensively studied for quantum computation~\cite{Loss1998,vandersypen2017}, but are also  a natural platform for implementing quantum simulations~\cite{Manousakis2002,Smirnov2007,Byrnes2008,Barthelemy2013,Gray2016}. The control over the electrical charge degree of freedom has facilitated the exploration of novel configurations such as effective attractive electron-electron interactions~\cite{Hamo2016}, collective Coulomb blockade~\cite{Hensgens2017}, and topological states~\cite{Kiczynski2022}. Coherent systems may be simulated when using the spin states of electrons in quantum dots, though experiments thus far have relied on gallium arsenide heterostructures ~\cite{Dehollain2020,Qiao2020,vanDiepen2021}, where the hyperfine interaction limits the spin coherence and therefore the complexity of simulations that can be performed. This bottleneck can be tackled by using group IV materials with nuclear spin-free isotopes. A natural candidate would be silicon, but this material comes with additional challenges due to the presence of valley states and a large effective electron mass~\cite{Zwanenburg2013}.

Hole quantum dots in planar Ge/SiGe heterostructures exhibit many favorable properties found in different quantum dot platforms~\cite{Scappucci2020}. Natural germanium has a high abundance of nuclear spin-free isotopes and can be isotopically purified~\cite{Sigillito2015}. Holes in germanium benefit from a low effective mass~\cite{Sammak2019,Lodari2019}, absence of valley degeneracies, ohmic contacts to metals~\cite{Hendrickx2018}, and strong spin-orbit coupling for all-electrical control~\cite{Watzinger2018,Hendrickx2020a}. Recent advances in heterostructure growth have resulted in stable, low-noise germanium devices~\cite{Lodari2021}. This has sparked rapid progress, with demonstrations of hole quantum dots~\cite{Hendrickx2018}, single hole qubits~\cite{Hendrickx2020a}, singlet-triplet (ST) qubits~\cite{Jirovec2021a}, two-qubit logic~\cite{Hendrickx2020b}, and a four qubit quantum processor~\cite{Hendrickx2021}.

 Here, we explore the prospects of hole quantum dots in Ge/SiGe for quantum simulation. We focus on the simulation of resonating valence bond (RVB) states, which are of fundamental relevance in chemistry~\cite{Pauling1931} and solid state physics~\cite{Anderson1987,Kivelson1987,Diep2013,Zhou2017} and have served as benchmark experiment in other platforms~\cite{Trebst2006,Nascimbene2012,Ma2011,Yang2021}. In our simulation, we probe RVB states in a square 2x2 configuration. First, we realize ST qubits for all nearest neighbours configurations. We then study the coherent evolution of four-spin states and demonstrate exchange control spanning an order of magnitude. Furthermore, we tune the system to probe valence bond resonances whose observed characteristics comply with predictions derived from the Heisenberg model. We finally demonstrate the preparation of $s$-wave and $d$-wave RVB states from spin-singlet states via adiabatic initialisation and tailored pulse sequences.

\begin{figure*}
\centering
\includegraphics[width=\textwidth]{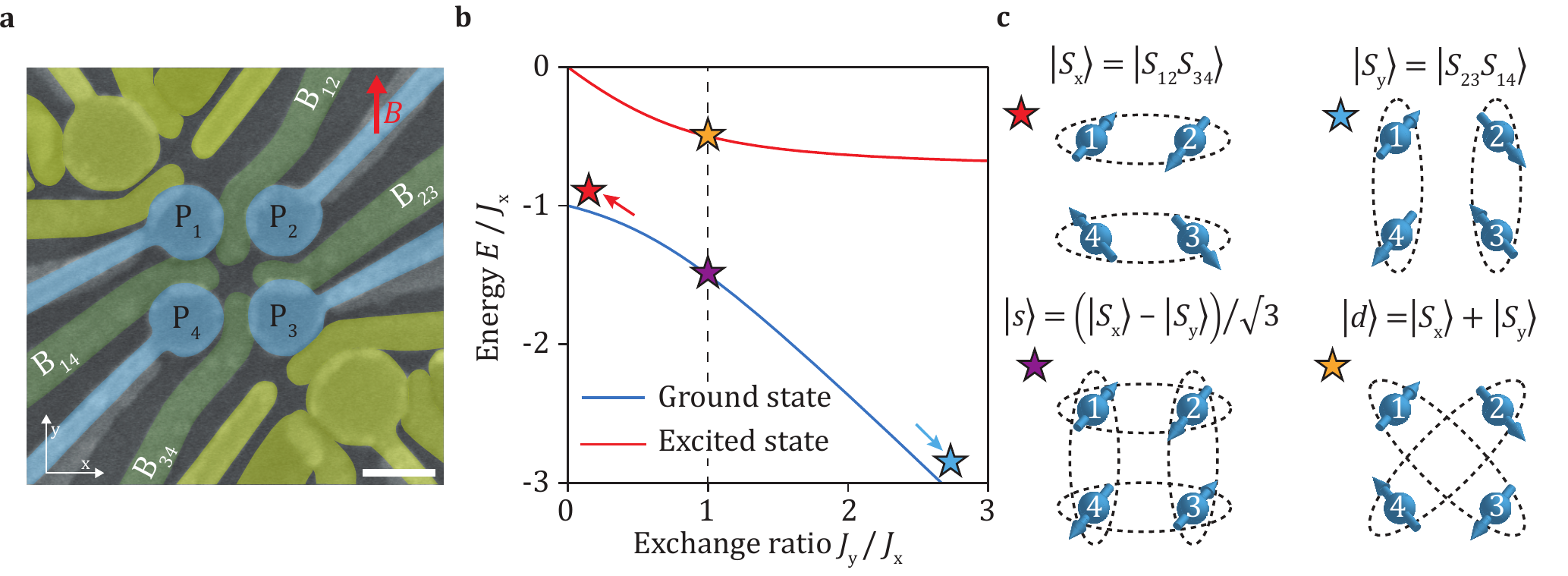}
\caption{\textbf{RVB states in a 2$\mathbf{\times}$2 quantum dot array.} \textbf{a}, False-coloured scanning electron micrograph of the Ge quantum dot array. Plunger and barrier gates are coloured in blue and green respectively, and the corresponding gate voltages applied on them are labelled. To achieve independent control of the quantum dot potentials and tunnel couplings, virtual plunger and barrier gate voltages are defined (see Supplementary Discussion~S1). Single hole transistors used as charge sensors are coloured in yellow. The scale bar corresponds to 100~nm. \textbf{b}, Energy diagram corresponding to the Hamiltonian $H_{S}$ and the stars denote the states as labeled in \textbf{c}. When the exchange interaction is dominated by horizontal (vertical) pairs, the ground state is $\ket{S_{\rm x}}$ ($\ket{S_{\rm y}}$), and in our experiments we use this configuration for initialization. Resonating valence bond states appear when $J_{\rm y}=J_{\rm x}$, the eigenstates are the ground state with $s$-wave symmetry and excited state with $d$-wave symmetry.}
\label{fig:Fig1}
\end{figure*}

\section*{RVB simulation in a quantum dot array with a square geometry}

% \blankfootnote{\Letter{~:~\href{mailto:C.C.Deprez@tudelft.nl}{C.C.Deprez@tudelft.nl},~\href{mailto:M.Veldhorst@tudelft.nl}{M.Veldhorst@tudelft.nl}}}

The experiments are based on a quantum dot array defined in a high-quality Ge/SiGe quantum well, as shown in Fig.~\ref{fig:Fig1}a~\cite{vanRiggelen2021,Hendrickx2021}. The array comprises four quantum dots and we obtain good control over the system, enabling to confine zero, one, or two holes in each quantum dot as required for the quantum simulation. The dynamics in resonating valence bands is governed by Heisenberg interactions. The spin states in germanium quantum dots, however, also experience Zeeman, spin-orbit and hyperfine interactions (see Supplementary Discussion S6). We therefore operate in small magnetic fields and acquire a detailed understanding of the system dynamics to apply tailored pulses. In the regime where Heisenberg interactions are dominating, the total spin is conserved. We can therefore study the subspaces of different total spin separately. The relevant subspace for the RVB physics is the zero total spin space spanned by the basis formed by the four spin states $\ket{S_{\rm x}}=\ket{S_{12}S_{34}}$ and $\frac{1}{\sqrt{3}}(\ket{T^+_{12}T^-_{34}}+
\ket{T^-_{12}T^+_{34}}-2\ket{T^0_{12}T^0_{34}})$, where $\ket{S_{\rm ij}}=\frac{1}{\sqrt{2}}(\ket{\uparrow_{\rm i} \downarrow_{\rm j}}- \ket{\downarrow_{\rm i} \uparrow_{\rm j}})$ and $\ket{T^0_{\rm ij}}=\frac{1}{\sqrt{2}}(\ket{\uparrow_{\rm i} \downarrow_{\rm j}}+ \ket{\downarrow_{\rm i} \uparrow_{\rm j}})$, $\ket{T^+_{\rm ij}}=\ket{\uparrow_{\rm i} \uparrow_{\rm j}}$, $\ket{T^-_{\rm ij}}=\ket{\downarrow_{\rm i} \downarrow_{\rm j}}$ denote the singlet and triplet states formed by the spins in the quantum dots i and j. In this basis, the Heisenberg Hamiltonian $H_{\rm J}$ reads:

\begin{equation}
H_{\rm J}(S_{\text{tot}}=0)\equiv H_{S} =
  \left( {\begin{array}{cc}
    -J_{\rm x}-\frac{J_{\rm y}}{4} & \frac{\sqrt{3}}{4}J_{\rm y} \\
  \frac{\sqrt{3}}{4}J_{\rm y} & -\frac{3}{4}J_{\rm y} \end{array} } \right),
\label{Hamiltonian_singlet_subspace}
\end{equation}

% \begin{equation}
% H_{S}=\frac{J_{\rm y}}{4}\begin{pmatrix}
% \frac{-4J_{\rm x}}{J_y}-1 & \sqrt{3\\
% \sqrt{3} & -3}

% \end{pmatrix}

% \end{equation}
where $J_{\rm x}=J_{12}+J_{34}$ and $J_{\rm y}=J_{14}+J_{23}$. Figures~\ref{fig:Fig1}.b-c show the eigenenergies and eigenstates of $H_{S}$ for different regimes of exchange interaction. When the exchange interaction is turned on in only one direction, $J_{\rm x}\gg J_{\rm y}$ or $J_{\rm x}\ll J_{\rm y}$, the system is equivalent to two uncoupled double quantum dots. The ground state is then a product of singlet states  $\ket{S_{ \rm x}}$ or $\ket{S_{\rm y}}=\ket{S_{14}S_{23}}$. However, when all exchanges are on and in particular when they are equal, $J_{\rm x}=J_{\rm y}$, the eigenstates are coherent superpositions of $\ket{S_{\rm x}}$ and $\ket{S_{\rm y}}$, which simulate the RVB state. In this regime, the ground state is the $s$-wave superposition state $\ket{s}=\frac{1}{\sqrt{3}}(\ket{S_{\rm x}}-\ket{S_{\rm y}})$ and the excited state is the $d$-wave superposition state $\ket{d}=\ket{S_{\rm x}}+\ket{S_{\rm y}}$. 

Fig.~\ref{fig:Fig1}.b shows that RVB states can be generated from uncoupled spin singlets by adiabatically equalizing the exchange couplings. Alternatively, if the exchange couplings are pulsed diabatically to equal values, valence bond resonances between $\ket{S_{ \rm x}}$ or $\ket{S_{\rm y}}$ states occur.

\section*{Singlet-Triplet oscillations in the four double quantum dots}

Probing the RVB physics relies on measuring the singlet probabilities in the (1,1) charge state~\cite{Nascimbene2012,vanDiepen2021}. We thus investigate ST oscillations within all nearest neighbour pairs.

\begin{figure}
\centering
\includegraphics[width=\columnwidth]{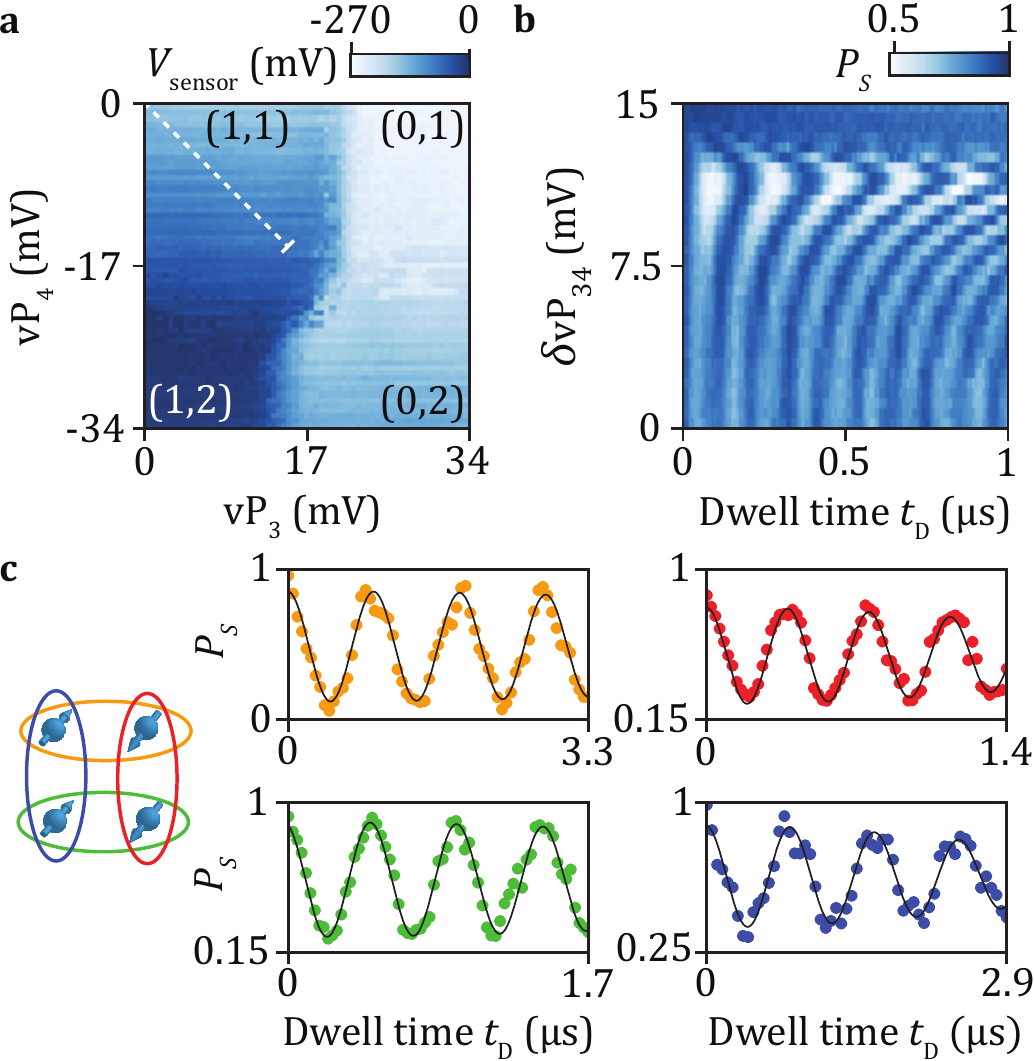}
\caption{\textbf{Singlet-triplet qubits on all nearest neighbour configurations.} \textbf{a}, Charge stability diagram of a double quantum dot (Q$_3$Q$_4$) in the few-hole regime. \textbf{b}, $S$-$T^-$ oscillations as a function of time and detuning $\delta\rm vP_{3,4}=0.5(vP_3-vP_4)$ varied along the dashed line in \textbf{a} . At larger magnetic fields, here $B$ = 3 mT, and limited tunnel couplings, we observe a minimum oscillation frequency due to the $S$-$T^-$ anticrossing. We therefore tune to smaller magnetic fields ($B$ = 1 mT) and larger tunnel couplings to operate away from this regime, where we observe, \textbf{c}, $S$-$T^{-}$ oscillations for all possible permutations of nearest neighbour quantum dot pairs. Black lines are fits of the data (see Supplementary Discussion S2).}
\label{fig:Fig2}
\end{figure}

To generate ST oscillations, we operate in a virtual gate landscape and apply pulses on the virtual plunger-gates $\rm vP_{\rm i}$ on each quantum dot pair according to the pulse sequence depicted in Fig.~\ref{fig:Fig2}.a ~\cite{Petta2005,Petta2010,Wu2014,Fogarty2018,Jirovec2021a,Jirovec2021b}. The double quantum dot system is initialized in a singlet (0,2) state. Then, the detuning between the quantum dots is varied by changing the virtual plunger gate voltages. The system is diabatically brought to a manipulation point in the (1,1) sector creating a coherent superposition of $\ket{S}$, $\ket{T^-}$ and $\ket{T^0}$~\cite{Petta2005,Petta2010,Wu2014,Fogarty2018,Jirovec2021a,Jirovec2021b}. After a dwell time $t_{\text{D}}$, the system is diabatically pulsed back to the (0,2) sector where the ST probabilities are determined via single-shot readout using (latched) Pauli-spin-blockade~\cite{Ono2002,Studenikin2012,HarveyCollard2018}.

Results of such experiments performed at $B=3$~mT with Q$_3$Q$_4$ pair are presented in Fig.~\ref{fig:Fig2}.b. Clear oscillations between the $\ket{S}$ and $\ket{T^{-}}$ state are observed over a large range of gate voltage. Importantly, using this method we find the $S$-$T^{-}$ anticrossing, which is the position where the frequency has a minimum. 
The observations of such oscillations, predominating over oscillations between $\ket{S}$ and $\ket{T^{0}}$ states,  agrees with recent investigations suggesting that $S$-$T^{-}$ oscillations dominate in germanium ST qubits placed in an in-plane $B$ field~\cite{Jirovec2021b}.

Fig.~\ref{fig:Fig2}.b also suggests that a (1,1)-singlet can be initialized from a (0,2)-singlet, while avoiding to pass the $S$-$T^-$ anticrossing, which would significantly improve the initialization. We achieve this by shifting the anticrossing towards the center of the (1,1) charge sector by decreasing the magnetic field to $B=1$~mT and increasing the tunnel coupling (Supplementary Figure~S1). Fig~\ref{fig:Fig2}.c demonstrates clear $S$-$T^-$ oscillations for all nearest neighbour configurations (see also Supplementary Figure~S2). Importantly, this also enables to determine the singlet/triplet states on two parallel quantum dot pairs by using sequential readout~\cite{Shulman2012}.

\section*{Tuning of individual exchanges using coherent oscillations}

The overlap of the $H_{S}$ eigenstates with $\ket{S_{\rm x}}$ and $\ket{S_{\rm y}}$ depends on $J_{\rm x}$ and $J_{\rm y}$~(see Supplementary Discussion S4). A quantitative comparison between experiments and theoretical expectations thus requires fine control over the exchange couplings. 

\begin{figure*}
	\centering
	\includegraphics[width=\textwidth]{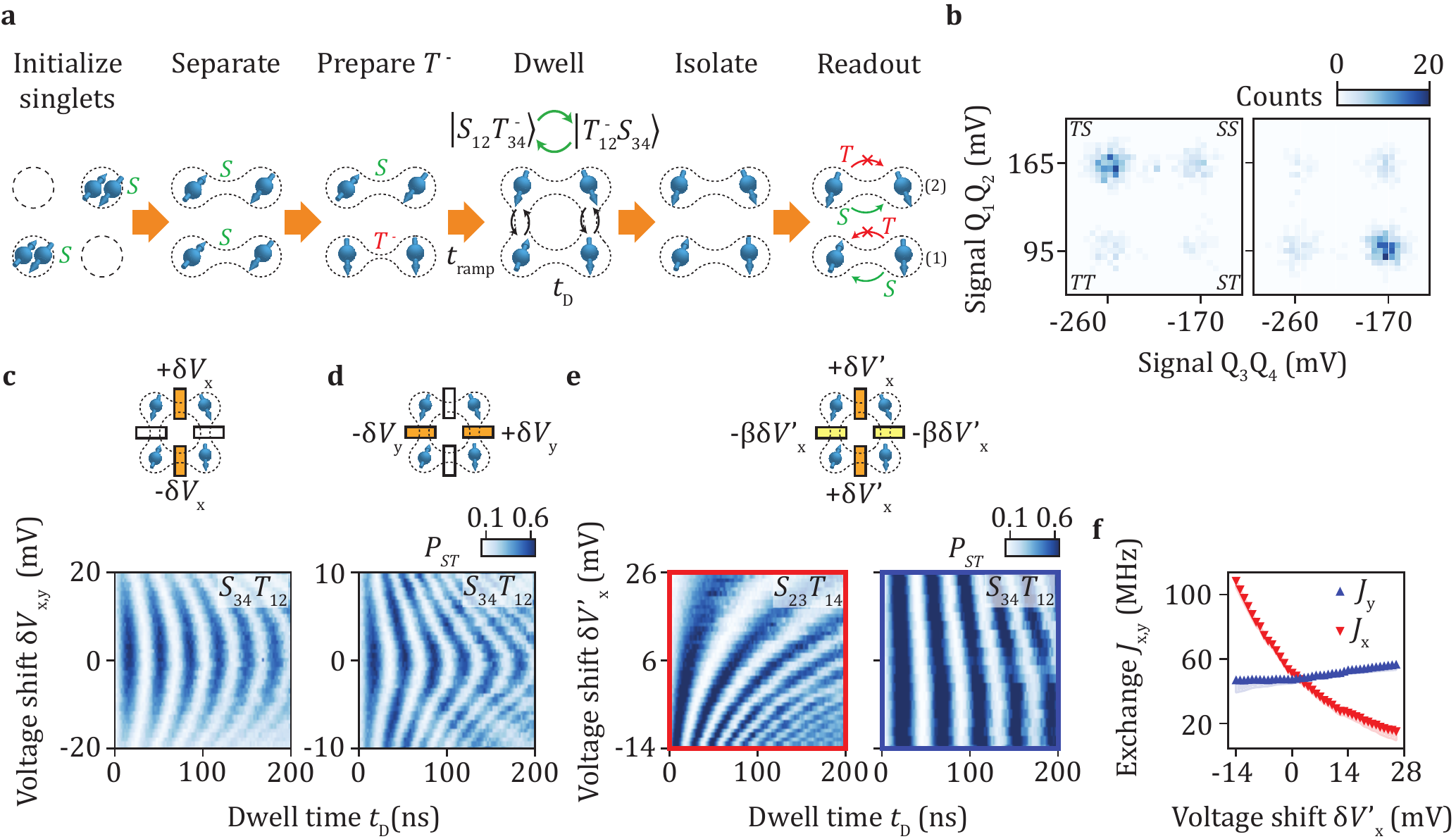}
		\caption{\textbf{Four-spin coherent singlet-triplet oscillations and exchange characterization.} \textbf{a}, Schematics of the pulse sequence used to measure four-spin ST oscillations from an initial $\ket{T_{34}S_{12}}$ state.  \textbf{b}, The 2D histograms of the sensor signals formed by sequential 500 single-shot readouts of Q$_3$Q$_4$ and Q$_1$Q$_2$ ST states. The left panel shows the initial state $T_{34}S_{12}$ at $t_{\rm D}=0$~ns. The right panel shows the state at  $t_{\rm D}=19$~ns corresponding to half an oscillation period. (Data corresponds to \textbf{c} for $\delta V_{\rm x}=0$.) \textbf{c}, Oscillations in ST probability  $P_{S_{34}T_{12}}$ as a function of gate voltage variation $\delta V_{\rm x}$. The amplitude of exchange pulses applied on the virtual barrier gates during the free evolution step are varied between measurements around a predetermined set of barrier gate voltages where $J_{\rm x} < J_{\rm y}$. The amplitude of voltage pulses on $\text{vB}_{12,34}$ are varied anti-symmetrically  while the amplitudes of the pulses on $\text{vB}_{23,14}$ are kept constant, as shown in the top illustration. The initial state is $\ket{T^{-}_{34}S_{12}}$. \textbf{d}, Similar experiment where oscillations in $P_{S_{34}T_{12}}$ are studied as a function of the gate voltage variation $\delta V_{\rm y}$. $\delta V_{\rm y}$ is the shift in the amplitudes of the exchange pulses applied anti-symmetrically on $\text{vB}_{23,14}$ (see top illustration). The initial state is $\ket{T^{-}_{34}S_{12}}$. \textbf{e}, Oscillations in $P_{S_{23}T_{14}}$ (left) and $P_{S_{34}T_{12}}$ (right) as functions of $t_{\rm D}$ and $\delta V^{\prime}_{\rm x}$. The amplitude of exchange pulses are varied symmetrically around the operation point ($\delta V^{\prime}_{\rm x}=0$~mV where $J_{\rm x}\simeq J_{\rm y}$) according to the top illustration ensuring that $J_{12}\simeq J_{34}$ and $J_{14}\simeq J_{23}$ along the full voltage range. The initial states are respectively a $\ket{T^{-}_{23}S_{14}}$ (left) and a $\ket{T^{-}_{34}S_{12}}$ state (right). \textbf{f}, Exchange couplings $J_{\rm x,y}$ extracted  by fitting the oscillations in  \textbf{e} with $A\cos(2\pi f_{ST} t_{\rm D}+\phi)\exp(-(t_{\rm D}/T_{\varphi})^2)+A_0$ as a function of gate voltage variation $\delta V^{\prime}_{\rm x}$. The oscillation frequencies of $P_{S_{23}T_{14}}$  ($P_{S_{34}T_{12}}$) corresponds to $J_{\rm x}/2$ ($J_{ \rm y}/2$). The shaded areas correspond to the estimated uncertainty on the exchange couplings derived based on assumptions discussed in Supplementary Discussion S5.}
	\label{fig:Fig3}
\end{figure*}

In this purpose, we focus on the evolution of coherent four-spin ST oscillations. These oscillations are induced using the experimental sequence depicted in Fig.~\ref{fig:Fig3}.a (see also Supplementary Figure S3). We turn off the vertical exchange coupling and initialize a $\ket{S_{\rm x}}$ or a $\ket{S_{\rm y}}$ state in the horizontal double quantum dots. We then rotate one of the singlet pairs to a triplet $\ket{T^-}$ state through coherent time evolution after pulsing to the $S$-$T^-$ anticrossing, creating a four-spin singlet-triplet product state (e.g. $\ket{T^{-}_{34}S_{12}}$ or $\ket{T^{-}_{23}S_{14}}$). All barrier gate voltages are  then diabatically pulsed to turn on all the exchange couplings leading to coherent evolution of the four-spin system. After a dwell time $t_{\rm D}$,  two pairs are isolated (not necessarily the initial ones) and their spin-states are read out sequentially, from which we can also deduce spin-correlation of opposite pairs, as was realized in linear arrays in GaAs~\cite{vanDiepen2021}.

The resonating valence bond requires equal couplings between all four quantum dots. In navigating to this point, we carefully develop a virtual landscape, keep control over all the individual exchange interactions. First, we separately equalize the horizontal ($J_{12}=J_{34}$) and vertical ($J_{14}=J_{23}$) exchange couplings. Then, we tune the vertical and horizontal exchanges to the same coupling strength. The Chevron patterns displayed in Fig.~\ref{fig:Fig3}.c-d are consistent with a Heisenberg Hamiltonian (see Supplementary Figures S4-S6) and the minima in the oscillation frequency mark the location of equal exchange coupling for horizontal ($J_{12}\simeq J_{34}\simeq J_{\rm x}/2$ for Fig.~\ref{fig:Fig3}.c) or vertical pairs  ($J_{14}\simeq J_{23}\simeq J_{\rm y}/2$ for Fig.~\ref{fig:Fig3}.d). Through an iterative process, we can find ranges of virtual gate voltages where $J_{12}\simeq J_{34}$ and $J_{23}\simeq J_{14}$.

We can now control the spin pairs simultaneously, while maintaining the exchange couplings in both the horizontal and vertical directions equal (see Supplementary Discussion S5), with \textit{a priori} $J_{\rm x}\neq J_{\rm y}$. In particular, virtual control enables to tune $J_{\rm x}$, while keeping $J_{\rm y}$ approximately constant, as illustrated in Fig.~\ref{fig:Fig3}.e~(see also Supplementary Discussion S1). Through readout of both pairs we obtain the oscillation frequency and with that the exchange interaction. Specifically, when a $\ket{T^-_{34}S_{12}}$  ($\ket{T^-_{23}S_{14}}$) state is initialized, the oscillation frequency is given by $f_{ST}=J_{\rm y}/2$  ($f_{ST}=J_{\rm x}/2$), highlighted in Fig.~\ref{fig:Fig3}.f (see Supplementary Discussion S5 for derivation). Although we observe a small dependence of $J_{\rm y}$, which varies from 46 MHz to 56 MHz, we are able to tune $J_{\rm x}$ by one order of magnitude from 15 MHz to 108 MHz. Clearly, the exchange interaction can be controlled and measured over a significant range and tuned to a regime where all couplings are equal (we obtain a precision of $\approx$~3~MHz, as discussed in Supplementary Discussion S5, mostly determined by drifts between experiments).

\section*{Valence bond resonances}
\begin{figure*}
	\centering
	\includegraphics[width=\textwidth]{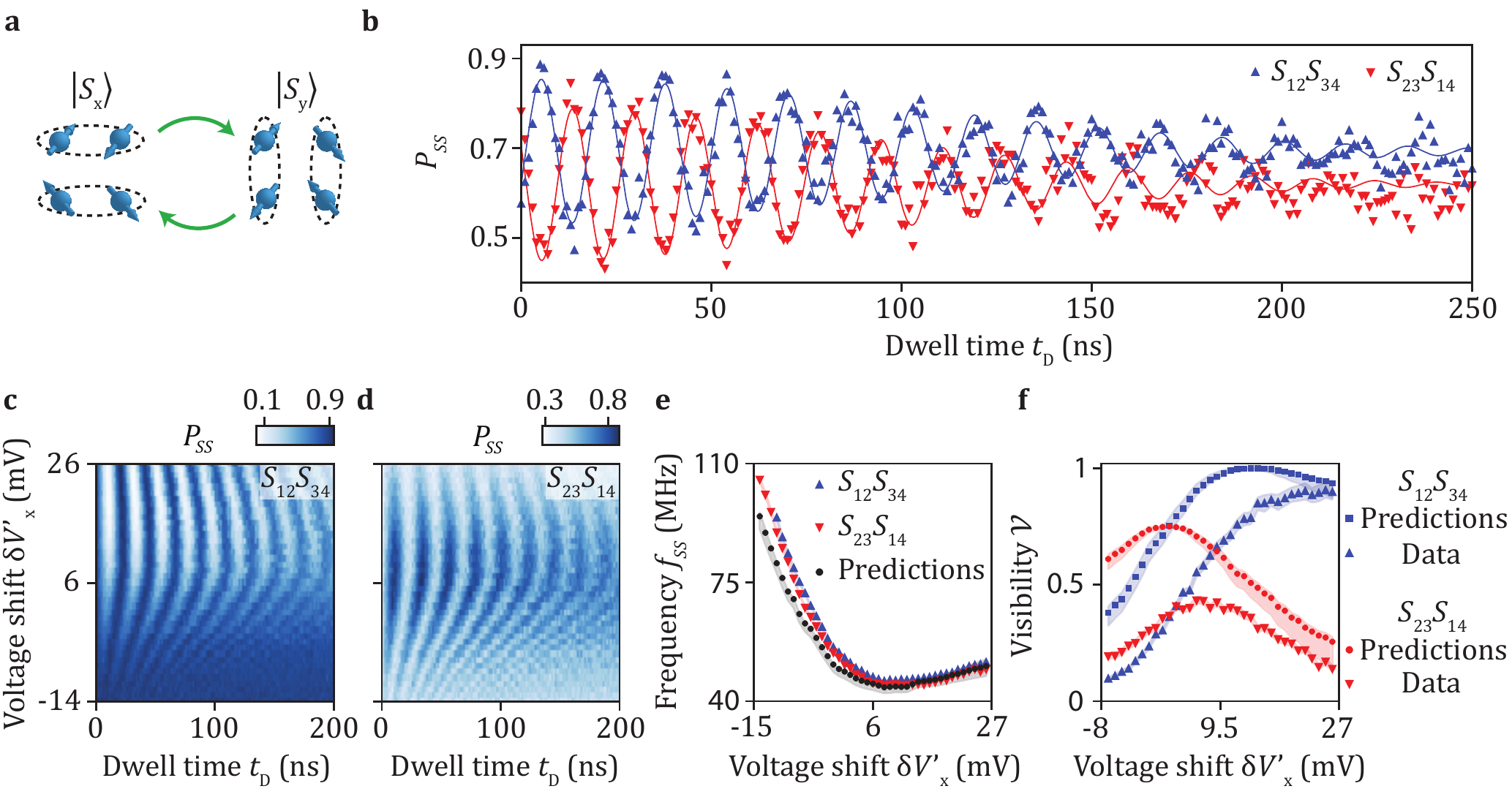}
		\caption{\textbf{Valence bond resonances.} \textbf{a}, Illustration showing valence bond resonances characterized by oscillations between the singlet product states $\ket{S_{\rm x}}$ and $\ket{S_{\rm y}}$. \textbf{b}, Probabilities of having horizontal singlet pairs $P_{S_{12}S_{34}}$ and vertical singlet pairs $P_{S_{23}S_{14}}$ as a function of dwell time $t_{\rm D}$. All the exchange couplings are tuned toward an identical value of $J_{\rm ij}\simeq25$~MHz. Lines are fits to the data with $P_{SS} = A\cos(2\pi f_{SS} t_{\rm D}+\phi)\exp(-(t_{\rm D}/T_{\varphi})^2)+A_0$ giving respectively  $T_{\varphi}=144$ and 130~ns for data corresponding to $P_{S_{12}S_{34}}$ and $P_{S_{23}S_{14}}$. The state is initialized as $\ket{S_{\rm y}}$. \textbf{c}, $P_{S_{12}S_{34}}$ and, \textbf{d}, $P_{S_{23}S_{14}}$ as a function of $t_{\rm D}$ and virtual barrier gate voltage variation $\delta V^{\prime}_{\rm x}$. The state is initialized as $\ket{S_{\rm x}}$. \textbf{e}, The oscillation frequency as a function of $\delta V^{\prime}_{\rm x}$. The blue (red) points are extracted from \textbf{c} (\textbf{d}). The black points are the theoretical predictions $f_{SS}=\sqrt{J_{\rm x}^2+J_{\rm y}^2-J_{\rm x}J_{\rm y}}$ computed using the exchanges $J_{\rm x}$ and $J_{\rm y}$ measured in Fig.~\ref{fig:Fig3}.e. \textbf{f}, Visibility as a function of gate voltage variation $\delta V^{\prime}_{\rm x}$. The triangles in blue (red) are extracted from \textbf{c} (\textbf{d}). The expected values are derived from equations~(S5) and (S7) using the measured exchanges. The shaded areas correspond to one standard deviation from the best fit for the experimental data, and for the theoretical data they correspond to the uncertainties on the amplitude and the frequency computed using the uncertainties on the exchange couplings values.}
	\label{fig:Fig4}
\end{figure*}

Valence band resonances can occur when all $J_{\rm ij}$ are equal. To experimentally assess this, we prepare $\ket{S_{\rm x}}$ or $\ket{S_{\rm y}}$, which are superposition states of $H_{S}$. We then pulse the exchange such that $J_{\rm x}\approx J_{\rm y}$. Fig.~\ref{fig:Fig4}.b shows the result of time evolution in this regime of equal exchange couplings. Since we start from a superposition of $H_{S}$, time evolution leads to coherent oscillations between $\ket{S_{\rm x}}$ or $\ket{S_{\rm y}}$, which we results in periodic swaps between the singlet states as depicted in Fig.~\ref{fig:Fig4}.a. In addition, we readout both in the horizontal and vertical configuration, and observe an anti-correlated signal, consistent with signatures of valence bond resonances~\cite{Kivelson1987,Nascimbene2012}. The observation of more than ten oscillations shows the relatively high level of coherence achieved during these experiments further confirmed by the characteristic dephasing time $T_{\varphi}\approx 130$~ns.

Fig.~\ref{fig:Fig4}.c-d show a more detailed measurement, which we can fit using $\frac{\mathcal{V}}{2}\cos(2\pi f_{SS} t_{\rm D}+\phi)\exp(-(t_{\rm D}/T_{\varphi})^2)+A_0$ to extract the evolution of the frequencies $f_{SS}$ and of the visibilities $\mathcal{V}$, plotted on Fig.~\ref{fig:Fig4}.e and Fig.~\ref{fig:Fig4}.f. We find a quantitative agreement between the measured frequencies and the theoretical expectation $f_{SS}=\sqrt{J_{\rm x}^2+J_{\rm y}^2-J_{\rm x}J_{\rm y}}$ despite deviations for the lowest values of $\delta V_{\rm x}^{\prime}$ that could result from the uncertainties in the exchange coupling. We also find a qualitative agreement for the visibilities $\mathcal{V}_{\rm x,y}$, though the measured $\mathcal{V}_{\rm x,y}$ remain lower, in particular when the exchange is larger. We ascribe this difference  to imperfect initialization and readout, leakage to states outside the relevant subspace, and insufficient diabaticity at large $J_{\rm x}$. However, the overall agreement confirms that the dynamics are governed by $H_{S}$ and that these oscillations are valence bond resonances. 

\section*{Preparation of resonating valence bond eigenstates}

\begin{figure}[h]
	\centering
	\includegraphics[width=\textwidth]{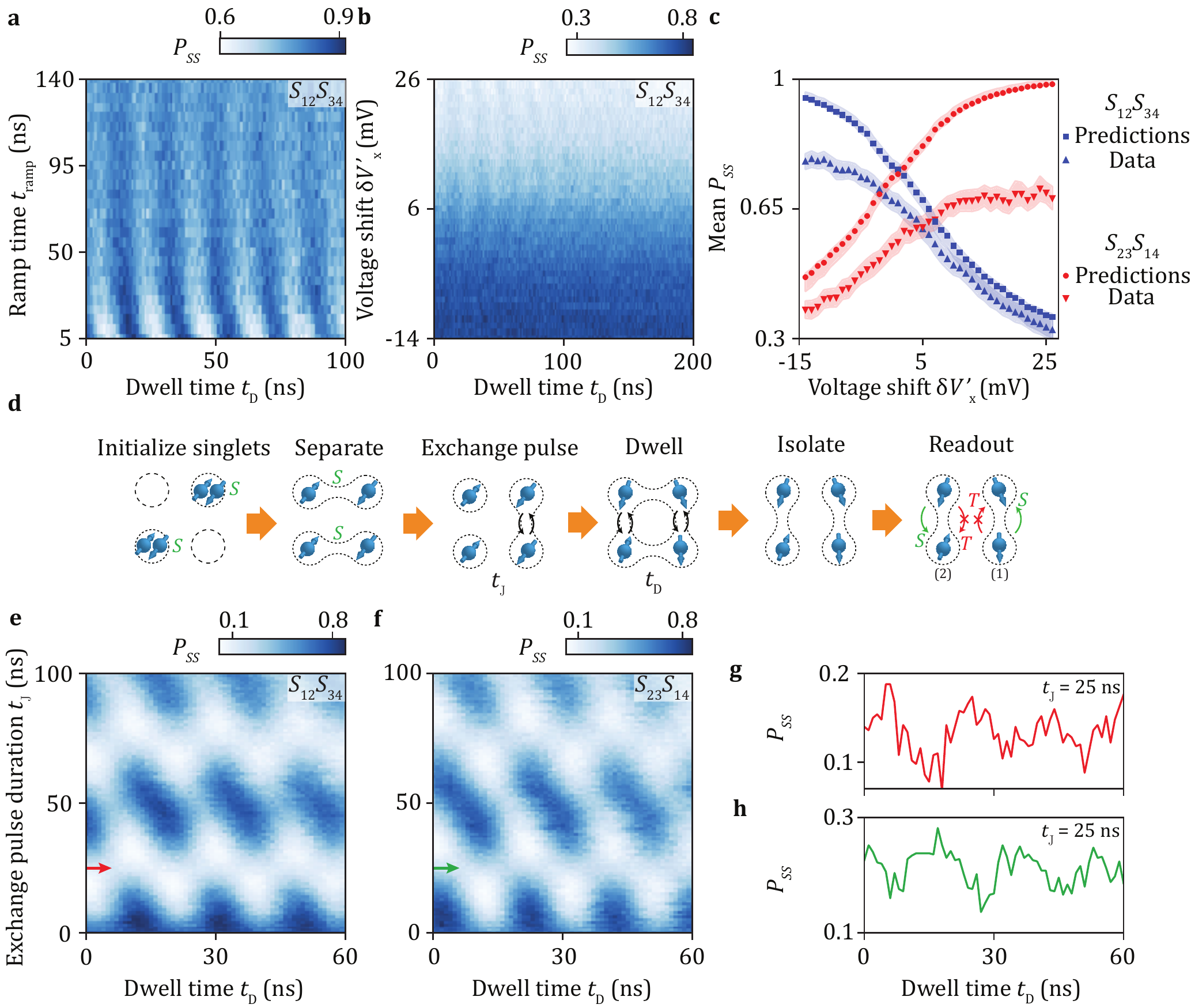}
		\caption{\textbf{Initialization of RVB eigenstates.} \textbf{a}, Evolution of RVB oscillations as a function of the time to set all exchanges equal ($J_{\rm ij}\simeq25$~MHz) (see Fig.~\ref{fig:Fig3}.a).  For $t_{\rm ramp}\gtrsim 140$~ns, the ground state with $s$-wave symmetry is adiabatically prepared.  \textbf{b}, Evolution of the singlet-singlet probability $P_{S_{12}S_{34}}$ with $\delta V^{\prime}_{\rm x}$ after adiabatic initialization of the ground state. \textbf{c}, Evolution of the mean singlet-singlet probability measured after adiabatic initialization of the ground state with $\delta V^{\prime}_{\rm x}$ for both readout directions. The experiments are compared to theoretical expectations using exchange coupling values extracted from global singlet-triplet oscillations (Supplementary Discussion S4). The shaded areas correspond to one standard deviation from the best fit for the experimental data, and for the theoretical data they correspond to the uncertainties on the amplitude and the frequency computed using the uncertainties on the exchange couplings values. \textbf{d}, Experimental sequence used to investigate the formation of the $d$-wave state. Before the free evolution step, one exchange pulses on $\rm vB_{23}$ is applied for a time $t_{\rm J}$. \textbf{e, f}, Evolution of singlet-singlet oscillations measured for different exchange pulse durations $t_{\rm J}$. The vanishing of oscillations at $t_{\rm J}\simeq25$~ns marks the formation of a $d$-wave state. \textbf{g, h}, Linecuts of \textbf{e} and \textbf{f} for $t_{\rm J}=25$~ns.}
	\label{fig:Fig5}
\end{figure}

Having observed valence bond resonances, we now focus on the preparation of eigenstates of $H_{S}$ which are the $\ket{s}$ and $\ket{d}$ RVB states. $\ket{s}$ is the ground state of $H_{S}$ when $J_{\rm x}=J_{\rm y}$, whereas $\ket{S_{\rm x}}$ and $\ket{S_{\rm y}}$ are the ground states when $J_{\rm x}\gg J_{\rm y}$ and $J_{\rm x}\ll J_{\rm y}$. Experimentally we therefore prepare $\ket{s}$ from $\ket{S_{\rm x}}$ or $\ket{S_{\rm y}}$ by adiabatically tuning the exchange interaction to equal values. Fig.~\ref{fig:Fig5}.a shows experiments where we control the ramp time $t_{\rm ramp}$ to tune to this regime and we observe a progressive vanishing of phase oscillations. For large $t_{\rm ramp}\gtrsim 140$~ns, the oscillations nearly disappear and the probability saturates to $P_{S_{12}S_{34}}\simeq 0.78$. Performing similar experiments starting from a $\ket{S_{\rm y}}$ state or measuring $P_{S_{23}S_{14}}$ leads to identical features with singlet-singlet probabilities saturating between 0.66 and 0.72 (see Supplementary Figure S9). These values are around the expectations ${\vert\braket{S_{\rm x,y}\vert s}\vert}^2=3/4$, expected when the $s$-wave state is prepared. 

We can now also prepare the ground state $H_{S}$ for arbitrary exchange values, by carefully tuning the ramp time ($t_{\rm ramp}=160$~ns in our experiments). Fig.~\ref{fig:Fig5}.b shows the evolution of $P_{S_{12}S_{34}}$ for different $\delta V^{\prime}_{\rm x}$. Since we prepare the ground state, coherent phase evolution results in a $P_{S_{12}S_{34}}$ that is virtually constant for any $\delta V^{\prime}_{\rm x}$ and only faint oscillations are observed. $P_{S_{12}S_{34}}$, however, is strongly dependent on $\delta V^{\prime}_{\rm x}$, as increasing $J_{\rm x}$ changes the ground state to $\ket{S_{\rm x}}$.

The measured $P_{S_{12}S_{34}}$ values can be compared with predictions using $J_{\rm x,y}$ values extracted from global singlet-triplet oscillations (see Supplementary Discussion S4). Fig.~\ref{fig:Fig5}.c shows that a good agreement exists between the theory and the experiments. The experimental $P_{S_{12}S_{34}}$ remains smaller than the theoretical predictions due to errors in the state preparation, initialisation, readout and leakage outside the global singlet subspace. Measuring $P_{S_{23}S_{14}}$ leads to a similar agreement although here imperfections have a larger impact.

\onecolumngrid\
\twocolumngrid

We prepare the $d$-wave state by including an additional operation where we exchange two neighbouring spins~\cite{Nascimbene2012}. This results in a transformation of neighbouring spin-spin correlations to diagonal correlations. We experimentally implement this step by adding, before the free evolution step, an exchange pulse of duration $t_{\rm J}$ during which only one exchange coupling is turned on (see Fig.~\ref{fig:Fig5}.d).

Fig.~\ref{fig:Fig5}.e-f shows $P_{S_{12}S_{34}}$ and $P_{S_{23}S_{14}}$ measured as functions of $t_{\rm D}$ and $t_{\rm J}$ in experiments where the system is initialized in $\ket{S_{\rm x}}$ and the exchange $J_{23}$ is pulsed. As a function of the exchange pulse duration, we observe a periodic vanishing of RVB oscillations (linecuts provided in Fig.~\ref{fig:Fig5}.g-h, imperfections in exchange control cause residual oscillations). Due to the exchange pulse, a periodic swapping of neighbouring spins occurs, and thus a periodic evolution between neighbouring spin-spin correlations and diagonal correlation. Thus the regime where the $d$-wave eigenstate is prepared is marked by the vanishing of RVB states. The mean of probabilities, $P_{S_{23}S_{14}}\simeq 0.21$ and $P_{S_{12}S_{34}}\simeq 0.13$, measured for $t_{\rm J}\simeq25~$ns are in the direction of theoretical expectations ${\vert\braket{S_{\rm x,y}\vert d}\vert}^2=1/4$. 

\section*{Conclusions}
In this work we demonstrated a coherent quantum simulation using germanium quantum dots. Clear evolution of resonating valence bond states, as well as the preparation of the $s$-wave and $d$-wave eigenstates. In addition, we have shown that we can control the exchange interaction over a significant range in a multi-spin setting.

The low-disorder and quantum coherence make germanium a compelling candidate for more advanced quantum simulations. Simultaneously, significant improvements have to be made in both the quantity and quality of the system. A significant improvement in the quantum coherence may be obtained by exploring sweet spots~\cite{Piot2022} and using purified germanium.

Controlling multi-spin states is also highly relevant in the context of quantum computation. The realization of exchange-coupled singlet-triplet qubits enables to implement fast two-qubit gates~\cite{Levy2002}, while the four-spin manifold provides means for decoherence-free subspaces~\cite{Lidar1998}. 

%Repeating similar analog simulations in larger germanium quantum dots arrays with different lattice geometries would offer a new tool to probe the physics of magnetic frustration, spin liquids and quantum phase transitions~\cite{Diep2013,Zhou2017}, potentially offering a quantum advantage with quantum dots.

Extensions of this work leveraging the full tunability of germanium quantum dots systems could provide new insights for extensive studies of strongly-correlated magnetic phases and  the associated quantum phase transitions. In particular, the implementation of similar simulations in triangular latices may allow to probe the transition between valence bond crystal phases and RVB spin liquid states~\cite{Diep2013} and to investigate the spin fractionalization in the latter case~\cite{Zhou2017}.  Similarly, such experiments could help to determine the debated nature of the phase at the $J_2=1/2~J_1$ point that emerge in square spin lattices with next-nearest neighbour antiferromagnetic exchange interactions~\cite{Diep2013}. Finally, our quantum dot platform allows to perform similar simulations with different charge configurations providing opportunities to explore the origins of superconductivity in doped cuprates~\cite{Anderson1987}.

\section*{Materials and Methods}

The device is made from a strained Ge/SiGe heterostructures grown by chemical vapour deposition. Starting from a natural Si wafer, a 1.6~$\mu$m thick relaxed Ge layer is grown, followed by a 1$~\mu$m reverse graded Si$_{1-x}$Ge$_x$ ($x$ going from 1 to 0.8) layer, a 500~nm relaxed Si$_{0.2}$Ge$_{0.8}$ layer, a 16~nm Ge quantum well under compressive stress, a 55~nm Si$_{0.2}$Ge$_{0.8}$ spacer layer and a $<1$~nm thick Si cap. The quantum well is contacted by aluminium ohmic contacts after a buffered oxide etch of the oxidized Si cap. The ohmics are isolated from the gates by a 10~nm ALD grown alumina layer. Two sets of Ti/Pd gates, separated by 7~nm of alumina, are deposited on top of the heterostructure to define the quantum dots. The potential of the quantum dots is tuned using the plunger gates (blue in Fig.~\ref{fig:Fig1}.a) while barrier gates are used to tune the tunnel couplings between the quantum dots (green). 

Virtual barrier and plunger gate voltages, defined as linear combinations of real gate voltages, are used to tune independently the potentials/couplings and compensate effects of cross-capacitances (see Supplementary Discussion S1) Experiments are performed in the 2$\times$2 array of quantum dots and change of the charge states in the array are measured using two single hole transistors (yellow in Fig.~\ref{fig:Fig1}.a) via rf-reflectometry. Further information regarding the experimental are provided in ref.~\cite{Hendrickx2021}.

For four-spin coherent oscillations, the spin-spin probabilities (or equivalently the spin-spin correlations) are determined by reading out sequentially the states of two parallel quantum dot pairs, either first Q$_3$Q$_4$ and then Q$_1$Q$_2$ or first Q$_2$Q$_3$ and then Q$_1$Q$_4$. While reading one pair, the second is stored deep in the (1,1) charge sector to prevent cross-talk between the measurements~\cite{Shulman2012,vanDiepen2021}. The state of each pairs is determined for each single shot-measurements by comparing the sensor signal to a predetermined threshold.

\section*{Acknowledgements}

We thank T.-K Hsiao, M. Russ, L. M. K. Vandersypen for their valuable advices and feedback. We also thank the other members of the Veldhorst and Vandersypen groups for stimulating discussions. We acknowledge O. Benningshof and R. Schouten for their technical support, and  S. G. J. Philips and S. L. de Snoo for their help on software development. 

We acknowledge support through an ERC Starting Grant and through an NWO projectruimte. Research was sponsored by the Army Research Office (ARO) and was accomplished under Grant No. W911NF-17-1-0274. The views and conclusions contained in this document are those of the authors and should not be interpreted as representing the official policies, either expressed or implied, of the Army Research Office (ARO), or the U.S. Government. The U.S. Government is authorized to reproduce and distribute reprints for Government purposes notwithstanding any copyright notation herein.

\section*{Author contributions}

C-A.W. and M.V. conceptualized the experiments. C-A.W., C.D. and H.T. conducted the measurements. C-A.W. and C.D. analysed the data. C.D. and M.V. wrote the paper with the inputs of all coauthors. W.I.L.L. fabricated the device. N.W.H. developed and designed the device. N.W.H. and W.I.L.L. built the experimental set-up. A.S. and G.S provided the heterostructure. M.V. supervised the project.

\section*{Competing interests}
The authors declare no competing interests. Correspondence should be sent to M. V. (M.Veldhorst@tudelft.nl).

\section*{Data availability}
All data underlying this study are available on a Zenodo repository at \url{https://doi.org/10.5281/zenodo.7016327}.

\bibliography{bib_RVB}

\clearpage

\end{document}

\title{Supplementary Information: Probing resonating valence bonds on a programmable germanium quantum simulator}

\author{Chien-An Wang}
\altaffiliation{These authors contributed equally}
\affiliation{QuTech and Kavli Institute of Nanoscience, Delft University of Technology, PO Box 5046, 2600 GA Delft, The Netherlands}
\author{Corentin Déprez}
\altaffiliation{These authors contributed equally}
\author{Hanifa Tidjani}
\affiliation{QuTech and Kavli Institute of Nanoscience, Delft University of Technology, PO Box 5046, 2600 GA Delft, The Netherlands}
\author{William I. L. Lawrie}
\affiliation{QuTech and Kavli Institute of Nanoscience, Delft University of Technology, PO Box 5046, 2600 GA Delft, The Netherlands}
\author{Nico W. Hendrickx}
\affiliation{QuTech and Kavli Institute of Nanoscience, Delft University of Technology, PO Box 5046, 2600 GA Delft, The Netherlands}
\author{Amir Sammak}
\affiliation{QuTech and Netherlands Organisation for Applied Scientific Research (TNO), Delft, The Netherlands}
\author{Giordano Scappucci}
\affiliation{QuTech and Kavli Institute of Nanoscience, Delft University of Technology, PO Box 5046, 2600 GA Delft, The Netherlands}
\author{Menno Veldhorst}
\affiliation{QuTech and Kavli Institute of Nanoscience, Delft University of Technology, PO Box 5046, 2600 GA Delft, The Netherlands}

\maketitle

\setcounter{table}{0}
\setcounter{figure}{0}
\setcounter{section}{0}
\makeatletter 
\renewcommand{\thefigure}{S\@arabic\c@figure}
\renewcommand{\thetable}{S\@arabic\c@table}
\renewcommand{\theequation}{S\@arabic\c@equation}
\renewcommand{\thesection}{S\@arabic\c@section}
\makeatother
\onecolumngrid
\begin{center}
\vspace{-1.5cm}
% \Letter{~:~\href{mailto:C.C.Deprez@tudelft.nl}{C.C.Deprez@tudelft.nl},~\href{mailto:M.Veldhorst@tudelft.nl}{M.Veldhorst@tudelft.nl}}
\end{center}
\vspace{0.5 cm}
This Supplementary Information includes :

\begin{itemize}
    \item Supplementary Discussions S1-S7
    \item Supplementary Figures S1-S9
    \item Supplementary Table S1
    \item Supplementary References S1-S5
\end{itemize}

\section{Virtual gate matrices}
\label{Section_virtual}
To independently control the chemical potentials of the quantum dots and  the sensors, we define the virtual plunger gates as linear combinations of the physical gate voltages. The virtual plunger-gate matrix is:

\begin{equation}
    \left( {\begin{array}{c}
 \rm P_{1} \\  \rm P_{2} \\ \rm  P_{3} \\  \rm P_{4} \\ \rm P_{\rm SHT_1} \\ \rm P_{SHT_2}
  \end{array} } \right)
  =
  \left( {\begin{array}{cccccc}
    1 & -0.28 & 0.03 & -0.2 & -0.14 & 0\\
    -0.26 & 1 & -0.27 & -0.01 & 0 & -0.02 \\
    0.02 & -0.2 & 1 & -0.29 & 0 & -0.08 \\
    -0.48 & -0.03 & -0.31 & 1 & 0 & 0 \\
    -0.12 & -0.03 & -0.01 & -0.02 & 1 & 0 \\
    0 & 0 & -0.12 & -0.03 & 0 & 1 
  \end{array} } \right) 
    \left( {\begin{array}{c}
  \rm vP_{1} \\  \rm  vP_{2} \\  \rm \rm vP_{3} \\  \rm vP_{4} \\\rm vP_{\rm SHT_1} \\\rm vP_{\rm SHT_2}
  \end{array} } \right),
\label{H_triplet_1}
\end{equation}
where $\rm P_{\rm SHT_{1,2}}$ are the voltages applied on the plunger gates of the sensors.

We also define virtual barrier gates enabling to have independent controls on each exchange meaning that a given exchange coupling can be tuned with virtually no change of the quantum dot potentials and of the other exchange couplings. We assume exponential models~\cite{Qiao2020} to describe the evolution of the exchange couplings with the virtual barrier gate voltages and find the corresponding coefficients by studying two-spin $S$-$T^-$ oscillations in the regime where spin pairs are decoupled  (either $J_{\rm x} \gg J_{\rm y}$ or $J_{\rm x} \ll J_{\rm y}$ ).  The virtual barrier gate matrix reads as:

\begin{equation}
  \left( {\begin{array}{c}
  \rm P_{1} \\  \rm P_{2} \\ \rm P_{3} \\ \rm  P_{4} \\ \rm  B_{12} \\ \rm  \rm B_{34} \\ \rm B_{23} \\ \rm B_{14} \\  \rm B_{SHT_1}
  \end{array} } \right) 
  =
  \left( {\begin{array}{cccc}
-0.564 & 0.042 &  0.076 & -0.181 \\
-1.296 &  0.492 & -1.212 &  0.713 \\
 0.048 & -0.554 & -0.16  & -0.062 \\
 0.65 &  -1.207 &  0.954 & -1.57 \\
 1    & -0.149 &  0.191 & -0.457 \\
 -0.227 &  1   &  -0.56 &   0.324 \\
 0.232 & -0.298 &  1   &  -0.228 \\
 -0.289 &  0.115 & -0.318 &  1 \\
 -0.012 &  0.015 & -0.05  &  0.011
  \end{array} } \right) 
  \left( {\begin{array}{c}
   \rm vB_{12} \\\rm vB_{34} \\\rm \rm vB_{23} \\ \rm vB_{14}
  \end{array} } \right),
\label{Virtual_matrix}
\end{equation}
with $\rm B_{\rm SHT_1}$ the voltages applied on the gate separating the dot Q$_1$ from the nearby sensor.

In practice, this method is not perfect when all the exchange interactions are turned on and some additional corrections are needed. For this reason, in four-spin experiments when $J_{\rm x}$ is varied by application of a $+\delta V^{\prime}_{\rm x}$ pulse on $\rm vB_{12}$ and $\rm vB_{34}$, we also apply a compensation pulse of $-0.18~\delta V^{\prime}_{\rm x} $ on $\rm vB_{23}$ and  $\rm vB_{14}$.

\section{$S$-$T^-$ oscillations in the double quantum dots at low magnetic field}

\begin{figure}[h!]
\centering
\includegraphics[width=\textwidth]{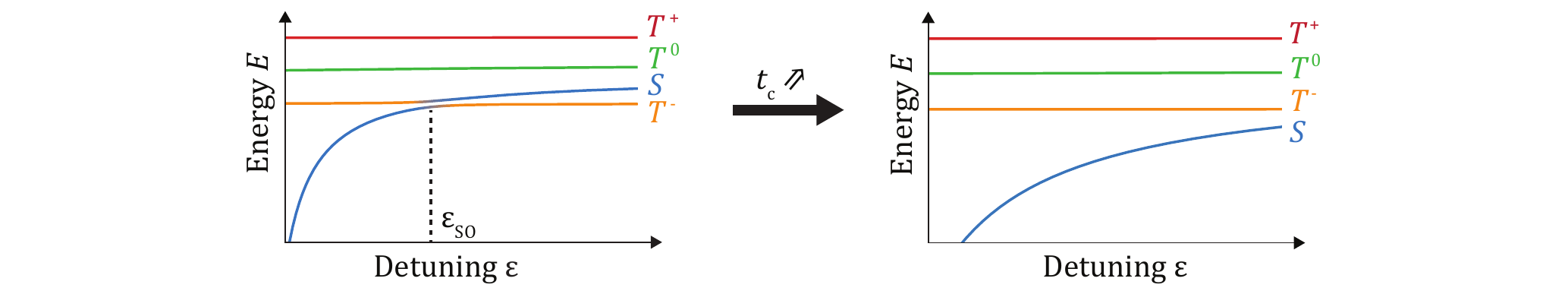}
\caption{ \textbf{Energy diagrams of a double quantum dot system at low fields.} At low tunnel couplings $t_{\rm c}$, the singlet $S$ energy state and the triplet $T^{-}$ anti-cross due to the spin-orbit interaction. Increasing $t_{\rm c}$, the energy of singlet state is lowered. At sufficiently large $t_{\rm c}$, the singlet state remains the ground state in the (1,1) charge state for any value of the detuning $\varepsilon$.}
\label{fig:Fig_ST_energy_diagram}
\end{figure}

In a double dot, the (1,1) singlet energy reads as $E_{S}\simeq\frac{\varepsilon}{2}-\sqrt{\frac{\varepsilon^2}{4}+2t_{\rm c}^2}$, with $t_{\rm c}$ the tunnel coupling between the quantum dots and $\varepsilon$ the detuning between the quantum dots (taken as zero at the (2,0)-(1,1) charge transition)~\cite{Jirovec2021a}. The energy of the triplet states are $E_{T^0}\simeq0$ and $E_{T^\pm}\simeq\pm\frac{\Sigma g}{2}\mu_B B$ with $\Sigma g$ the $g$-factor sum. The corresponding energy diagram is sketched in Fig.~\ref{fig:Fig_ST_energy_diagram}. 

In the (1,1) charge sector, the ground state is the singlet $\ket{S}$. Above a given value of detuning $\varepsilon_{\rm SO}$, the $\ket{T^-}$ state becomes the ground state. Consequently there is an anti-crossing between the $\ket{S}$ and $\ket{T^-}$ due to the spin-orbit interaction. In order to maintain the singlet ground state in the (1,1) charge sector, one can decrease the magnetic field $B$ or increase the tunnel coupling $t_{\rm c}$.

At $B=1$~mT, the ground state is the singlet state for Q$_1$Q$_2$ and Q$_1$Q$_4$ pairs in the whole (1,1) charge sector. It gives more freedom for the singlet-(1,1) initialization with these two quantum dot pairs. For the Q$_2$Q$_3$ and Q$_3$Q$_4$, there is still a $S$-$T^-$ anticrossing that appears at finite detuning but sufficiently far from the charge transition line to reduce charge noise effects detrimental for four spin experiments.

To observe high visibility ST oscillations, pulses on the virtual barrier voltages are applied to reduce $t_{\rm c}$ while going from the initialization to the manipulation point. This configuration offers more flexibility to initialize $\ket{S_{\rm x}}$ or $\ket{S_{\rm y}}$ states.

\begin{figure}[h!]
\centering
\includegraphics[width=\textwidth]{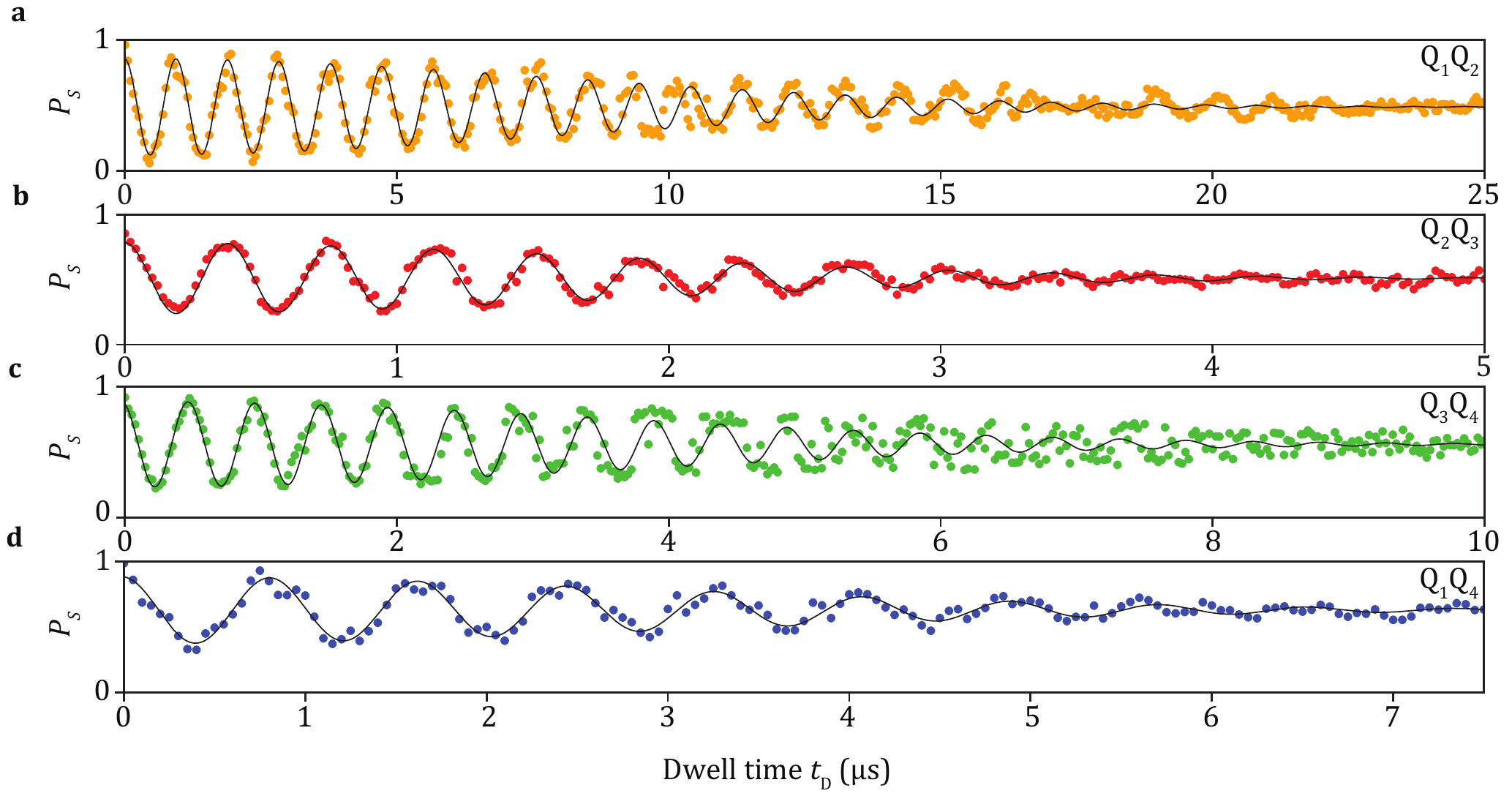}
\caption{\textbf{Singlet-Triplet oscillations observed with each double quantum dot.} Same measurements than ones showed in Fig.~2.c. Data (points) are fitted with $A\cos(2\pi f t_{\rm D}+\phi)\exp(-(t/T_{\varphi})^2)+A_0$ to extract $T_{\varphi}$ and $f$.}
\label{fig:Fig_Supp_ST_pairs}
\end{figure}

Fig.~\ref{fig:Fig_Supp_ST_pairs} presents the ST oscillations of Fig.~2.c over larger ranges of dwell time $t_{\rm D}$. By fitting the data, we extract the characteristic dephasing times $T_{\varphi}$ and the frequency of the oscillations $f$ for each pair (Table~\ref{table:Table_ST_pairs}). We note that there are large variations of both $T_{\varphi}$ and $f$. The variation of $f$ can be explained by differences in the strengths of the tunnel couplings $t_{\rm c}$, the  differences in the $g$-factor and  in the amplitudes of the barrier voltage pulses. They lead to variations of the energy splitting between the $\ket{T^{-}}$ and $\ket{S}$ states. The variations of $T_{\varphi}$ can result from  different effects like residual exchange interactions with the other quantum dots or leakages to the $\ket{T^{0}}$ states. The lower coherence of Q$_2$Q$_3$ and  Q$_1$Q$_4$ pairs compared to that of  Q$_1$Q$_2$ and  Q$_3$Q$_4$ could also result from the field orientation: the spin life time is indeed reduced when the spin-orbit field is oriented perpendicular to the external magnetic field~\cite{Danon2009,Hendrickx2021}.  

\begin{table}[h!]
\begin{center}
\begin{tabular}{|c|c|c|} 
 \hline
 Dot Pair & Frequency (MHz) & $T_{\varphi}$ ($\mu$s) \\ [0.5ex] 
 \hline\hline
 Q$_1$Q$_2$ & $1.056\pm0.001$ & $11.2\pm0.4$ \\ 
 \hline
 Q$_2$Q$_3$ & $2.636\pm0.005$ & $2.5\pm0.1$ \\
 \hline
 Q$_3$Q$_4$ & $2.043\pm0.004$ & $5.1\pm0.3$\\
 \hline
 Q$_1$Q$_4$ & $1.223\pm0.004$ & $4.2\pm0.2$\\
 \hline
\end{tabular}
\end{center}
\caption{\textbf{Characteristics of singlet-triplet oscillations of individual dot pairs.} The uncertainties correspond to one standard deviation from the best fits.}
\label{table:Table_ST_pairs}
\end{table}

\section{Pulse sequences to generate global ST oscillations}
\begin{figure}[h!]
	\centering
	\includegraphics[width=\textwidth]{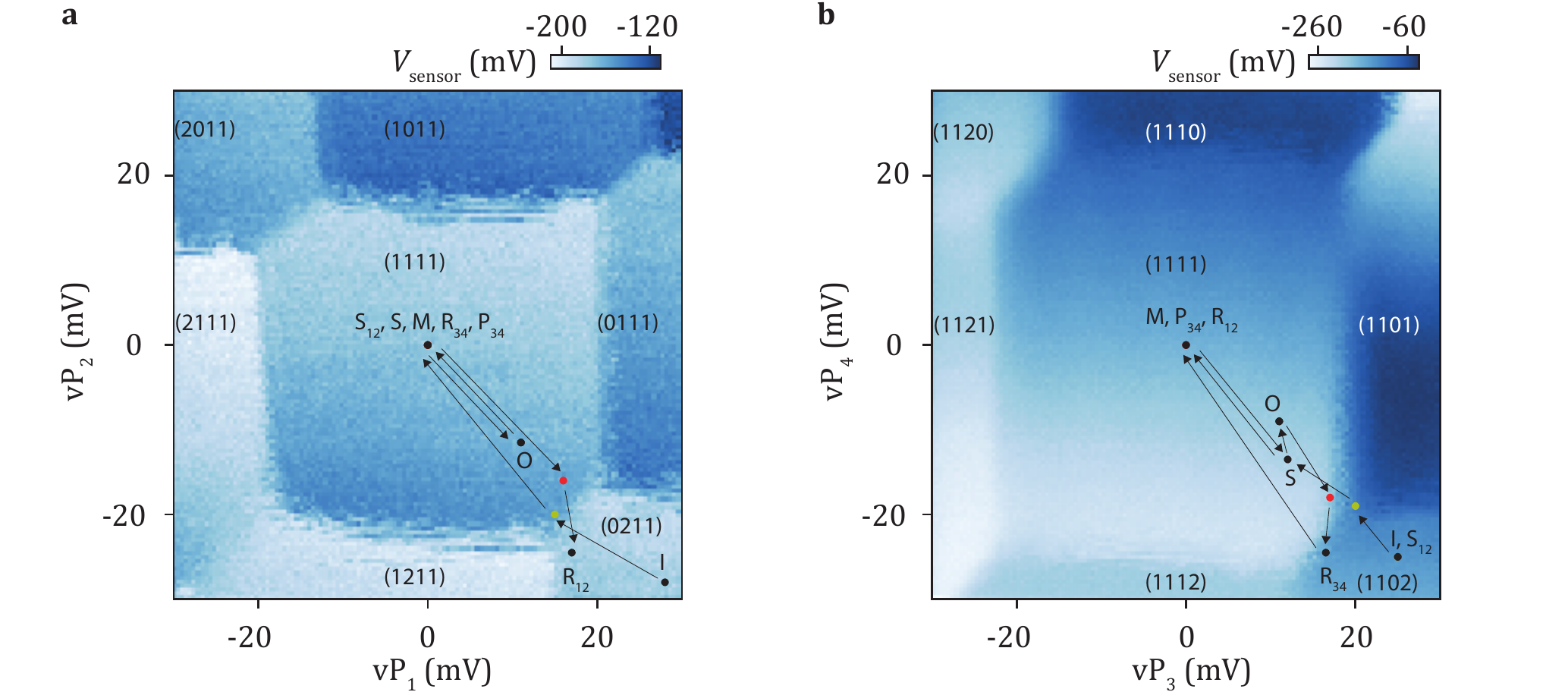}
		\caption{\textbf{Charge Stability Diagram (CSD) and the projected pulse sequences for experiments of four-spin singlet-triplet oscillations.}  \textbf{a}, CSD of Q$_1$Q$_2$ pair as a function of  $\rm vP_{1}$ and $\rm vP_{2}$. \textbf{b}, CSD of Q$_3$Q$_4$ pair as a function of  $\rm vP_{3}$ and $\rm vP_{4}$. When the CSD of one pair is measured, the other pair is in  the (1,1) charge state.  This makes the initialization point I to appear in the (0211) and  the (1102) charge occupation whereas the actual occupation at I is (0202). The plunger gate voltages used in the pulse sequence are indicated by black points. The red and yellow points are additional steps between separation process and readout process that allows to make sure that the system goes through the charge transition line.}
	\label{fig:Fig_Sup_CSD}
\end{figure}

The pulse sequence for generating and measuring global ST oscillations (Fig.~3) is plotted on charge stability diagrams in Fig.~\ref{fig:Fig_Sup_CSD}. The black points indicate the virtual plunger gate voltages at each stage of the sequence depicted in Fig.~3.a. 

The spins are initialized at (0,2,0,2) as marked by the label I. The pair Q$_1$Q$_2$ is separated first by pulsing to the point $S_{12}$ while the pair Q$_3$Q$_4$ stays in (0,2), and then the pair Q$_3$Q$_4$ is separated by pulsing to the point S. The system is initialized as $\ket{S_{12}S_{34}}$. Until this point, the two pairs are decoupled by application of sufficiently large voltages on $\rm vB_{23}$ and $\rm vB_{14}$. 

Next, we diabatically pulse the virtual plungers and barriers to the point M in Fig.~\ref{fig:Fig_Sup_CSD}, wait for $t_{\pi}=300$~ns and then pulse back to the point S. The barrier pulse is optimized to have the minimal level spacing between the $\ket{S}$ and $\ket{T^-}$ state and the largest contrast of $S$-$T^-$ oscillation at the point M. The waiting time $t_{\pi}$ enables the rotation from the $\ket{S_{34}}$ to $\ket{T^-_{34}}$ state.

At that point, the system is prepared in a $\ket{S_{12}T^-_{34}}$ state. After this step, we first pulse the plunger gates to the point O and then the barrier gates (with a ramp time $\rm t_{ramp}$). It allows to turn on all the exchange interactions. We wait in this setting for a dwell time $\rm t_{\rm D}$.

After the free evolution, the pairs are isolated by increasing the virtual barrier voltages. For the horizontal readout, the virtual barriers $\rm vB_{23}$ and $\rm vB_{14}$ are pulsed diabatically to isolate pair Q$_1$Q$_2$ and pair Q$_3$Q$_4$. For the vertical readout, the barriers $\rm vB_{12}$ and $\rm vB_{34}$ are pulsed diabatically to form isolated pairs Q$_2$Q$_3$ and Q$_1$Q$_4$. The isolation continues until the readout of both pair states is performed.

The horizontal readout is done by first reading out pair Q$_3$Q$_4$ and then pair Q$_1$Q$_2$. The pair Q$_3$Q$_4$ is pulsed to the point R$_{34}$ to perform Pauli spin blockade readout while pair Q$_1$Q$_2$ stays deep in (1,1). After integrating the charge signal for 3~$\mu$s, the pair Q$_3$Q$_4$ is brought back to the point P$_{34}$  deep in (1,1) sector such that its charge state does not affect subsequent readout on the pair Q$_1$Q$_2$. Afterwards, the pair Q$_1$Q$_2$ is pulsed to the point R$_{12}$ and the charge signal is integrated for 3~$\mu$s. 

The pulse sequences for different readout or initialization directions have similar structures.

\section{Four-spin coherent oscillations in the global singlet subspace}
\subsection{Theoretical model}
The Heisenberg Hamiltonian $H_{S}$ in the global singlet subspace can be written, up to an overall energy shift, as:
\begin{equation}
H_{S} =
(-\frac{1}{2} J_{\rm x} + \frac{1}{4} J_{\rm y} ) \sigma_{\rm z} +
\frac{\sqrt{3}}{4} J_{\rm y} \sigma_{\rm x} = h_0 \cos{\theta} \sigma_{\rm z} + h_0 \sin{\theta} \sigma_{\rm x},
\label{Hamiltonian_singlet_subspace_suppl}
\end{equation}

where  $\cos{\theta}=\frac{-2J_{\rm x}+J_{\rm y}}{2\sqrt{J_{\rm x}^2-J_{\rm x}J_{\rm y} + J_{\rm y}^2}}$, $\sin{\theta}=\frac{\sqrt{3}J_{\rm y}}{2\sqrt{J_{\rm x}^2-J_{\rm x}J_{\rm y} + J_{\rm y}^2}}$, $h_0=\frac{1}{2} \sqrt{J_{\rm x}^2-J_{\rm x}J_{\rm y} + J_{\rm y}^2} $, and $\sigma_{\rm x,z}$ are the Pauli matrices. Here we denote the basis states $\{ \ket{0}, \ket{1} \}
\equiv
\{    \ket{S_{12}S_{34}}, \frac{1}{\sqrt{3}}(\ket{T^+_{12}T^-_{34}}+
\ket{T^-_{12}T^+_{34}}-2\ket{T^0_{12}T^0_{34}}) \}
\equiv
\{ \begin{pmatrix}1 \\ 0 \end{pmatrix} , \begin{pmatrix}0 \\ 1 \end{pmatrix}  \}$ .

The eigenenergies are  $E_{\rm g} = -h_0 $ and $E_{\rm e} = h_0 $. The eigenstates are:

\begin{equation}
\begin{cases}
\ket{g} = \begin{pmatrix}- \sin{\frac{\theta}{2}} \\ \cos{\frac{\theta}{2}} \end{pmatrix}  \\
\ket{e} =   \begin{pmatrix}  \cos{\frac{\theta}{2}} \\ \sin{\frac{\theta}{2}} \end{pmatrix}
\end{cases}.
\label{States_singlet}
\end{equation}

In the singlet-singlet oscillation experiments, the state is initialized in a singlet-singlet state $\ket{S_{12}S_{34}}=\begin{pmatrix}1 \\ 0 \end{pmatrix}$ which can be written as $-\sin{\frac{\theta}{2}} \ket{g} + \cos{\frac{\theta}{2}} \ket{e}$. After the free evolution this state becomes, up to a phase factor: 

\begin{equation*}
\ket{\psi(t)} = 
-\sin{\frac{\theta}{2}} \ket{g} + \cos{\frac{\theta}{2}} e^{\rm -i \omega_{\rm eg} t} \ket{e} = 
\begin{pmatrix}
\sin^2{(\frac{\theta}{2})} + \cos^2{(\frac{\theta}{2})} e^{\rm -i \omega_{\rm eg} t}  \\
\sin{(\frac{\theta}{2})} \cos{(\frac{\theta}{2})} (-1 + e^{\rm -i \omega_{\rm eg} t} )  
\end{pmatrix} 
= 
e^{\rm -i \omega_{\rm eg} t/2}
\begin{pmatrix}
\cos{\frac{\omega_{\rm eg} t}{2}} - i \cos{\theta} \sin{\frac{\omega_{\rm eg} t}{2}}   \\
 - i \sin{\theta} \sin{\frac{\omega_{\rm eg} t}{2}}
\end{pmatrix} ,
\label{Evolve_singlet}
\end{equation*}

where $\omega_{\rm eg} \equiv \omega_{\rm e} - \omega_{\rm g} = \frac{1}{\hbar} \sqrt{J_{\rm x}^2 - J_{\rm x}J_{\rm y} + J_{\rm y}^2} $ is the frequency of the singlet oscillations.

The probability of being in the state $\ket{0}$ is $P_{S_{34}S_{12}}(t) = \vert\braket{S_{12}S_{34}\vert\psi(t)}\vert^2= \frac{1}{2 } (1+\cos^2{\theta} + \sin^2{\theta} \cos{\omega_{\rm eg} t})$. The visibility of the oscillations is then:
\begin{equation}
\mathcal{V}_{\rm x}=P^{\rm max}_{S_{34}S_{12}}-P^{\rm min}_{S_{34}S_{12}} = \sin^2{\theta} = \frac{3J_{\rm y}^2}{4 (J_{\rm x}^2 - J_{\rm x}J_{\rm y} + J_{\rm y}^2)}.
\label{Visibility_SS_osc_x}
\end{equation}

To describe the readout in the y direction, we use the basis $\{ \ket{0_{\rm y}}, \ket{1_{\rm y}} \}=\{    \ket{S_{14}S_{23}}, \frac{1}{\sqrt{3}}(\ket{T^+_{14}T^-_{23}}+
\ket{T^-_{14}T^+_{23}}-2\ket{T^0_{14}T^0_{23}}) \}$. The original basis can be re-written in terms of the new basis as:

\begin{equation}
\begin{cases}
\ket{0} = -\frac{1}{2}\ket{0_{\rm y}} - \frac{\sqrt{3}}{2} \ket{1_{\rm y}} \\
\ket{1} = \frac{\sqrt{3}}{2}\ket{0_{\rm y}} - \frac{1}{2} \ket{1_{\rm y}}
\end{cases} .
\label{Basis_transform_singlet}
\end{equation}

Therefore, $P_{S_{23}S_{14}}(t) 
= |\braket{0_{\rm y} | \psi(t)}|^2  = \frac{1}{4}(1 + \sin^2{\theta}-\sqrt{3}\sin{\theta}\cos{\theta})
+
\frac{1}{4}(- \sin^2{\theta}+\sqrt{3}\sin{\theta}\cos{\theta}) \cos{\omega_{\rm eg} t}=\frac{1}{4}(1+(\sin^2{\theta}-\sqrt{3}\sin{\theta}\cos{\theta})\left(1-\cos{\omega_{\rm eg} t})\right)
$. 
The visibility is then:

\begin{equation}
\mathcal{V_{\rm y}}=\frac{1}{2}(\sin^2{\theta}-\sqrt{3}\sin{\theta}\cos{\theta})=\frac{3J_{\rm x} J_{\rm y}}{4 (J_{\rm x}^2 - J_{\rm x}J_{\rm y} + J_{\rm y}^2)}.
\label{Visibility_SS_osc_y}
\end{equation}

We note that $\sin^2{\theta}-\sqrt{3}\sin{\theta}\cos{\theta}=\frac{6 J_{\rm y} J_{\rm x}}{4(J_{\rm x}^2 - J_{\rm x}J_{\rm y} + J_{\rm y}^2)}>0$ and thus $P_{S_{34}S_{12}}(t)$ and  $P_{S_{23}S_{14}}(t)$ oscillate in phase opposition. There are periodic swaps between $\ket{S_{\rm x}}$ and $\ket{S_{\rm y}}$ which are the resonating valence bond oscillations as shown in Fig.~4.

\subsection{Singlet probabilities of s-wave and d-wave states}

To prepare the $s$-wave and $d$-wave states, the exchanges are set to be equal. It corresponds to the Hamiltonian of equation~\ref{Hamiltonian_singlet_subspace_suppl} with  $\theta=120^{\circ}$. The $s$-wave state is the ground state and reads $\ket{s} = \ket{g}=(-\frac{\sqrt{3}}{2},\frac{1}{2})$. The singlet-singlet probability in both x and y directions for this state are $P_{S_{34}S_{12}} = P_{ S_{23}S_{14}} = \frac{3}{4}$. The d-wave state is the excited state and reads $\ket{d} = \ket{e}=(\frac{1}{2}, \frac{\sqrt{3}}{2})$. The singlet-singlet probabilities for this state are $P_{S_{34}S_{12}} = P_{S_{23}S_{14}} = \frac{1}{4}$.

When the exchanges are different, the equation~\ref{States_singlet} gives the ground state singlet-singlet readout probability $P_{S_{34}S_{12}} = \sin^2{\frac{\theta}{2}} = \frac{1-\cos{\theta}}{2}=  \frac{1}{2} - \frac{-2J_{\rm x}+J_{\rm y}}{4\sqrt{J_{\rm x}^2-J_{\rm x}J_{\rm y} + J_{\rm y}^2}}$ and 
$P_{S_{23}S_{14}} = 
(\frac{1}{2} \sin{\frac{\theta}{2}} + \frac{\sqrt{3}}{2} \cos{\frac{\theta}{2}})^2 = 
\frac{1}{2} +  \frac{1}{4} \cos{\theta} +  \frac{\sqrt{3}}{4} \sin{\theta} = 
\frac{1}{2} + \frac{-J_{\rm x}+2J_{\rm y}}{4\sqrt{J_{\rm x}^2-J_{\rm x}J_{\rm y} + J_{\rm y}^2}}$. These formula are used in Fig.~5.c.

\section{Four-spin coherent oscillations in the global triplet subspace}
\label{Section_Four_Spin_ST_osc}
\subsection{Theoretical model}
In this section, we derive the theoretical results used to infer the exchange coupling $J_{\rm x,y}$ from the four-spin singlet-triplet oscillations. In our experiments, we operated in the  $m_{S}=-1$ global triplet subspace spanned by a natural basis $\{ \ket{S_{12}T_{34}^{-}} , \ket{T_{12}^{-}S_{34}}, \frac{1}{\sqrt{2}}( \ket{T_{12}^{0}T_{34}^{-}} - \ket{T_{12}^{-}T_{34}^{0}})\}$. Considering only Heisenberg exchange interactions, the Hamiltonian can be written as:

\begin{equation}
H_{T} =
  \left( {\begin{array}{ccc}
    -J_{\rm 12}-\frac{J_{\rm 23}+J_{\rm 14}}{4} & -\frac{J_{\rm 23}+J_{\rm 14}}{4} & -\frac{J_{\rm 23}-J_{\rm 14}}{2\sqrt{2}} \\
  -\frac{J_{\rm 34}+J_{\rm 14}}{4} & -J_{\rm 34}-\frac{J_{\rm 23}+J_{\rm 14}}{4} & -\frac{J_{\rm 23}-J_{\rm 14}}{2\sqrt{2}} \\
  -\frac{J_{\rm 23}-J_{\rm 14}}{2\sqrt{2}} & -\frac{J_{\rm 23}-J_{\rm 14}}{2\sqrt{2}} & -\frac{J_{\rm 23}+J_{\rm 14}}{2} 
  \end{array} } \right)=
    \left( {\begin{array}{ccc}
    -\frac{J_{\rm x}+\delta_{\rm x}}{2}-\frac{J_{\rm y}}{4} & -\frac{J_{\rm y}}{4} & -\frac{\delta_{\rm y}}{2\sqrt{2}} \\
  -\frac{J_{\rm y}}{4} & -\frac{J_{\rm x}-\delta_{\rm x}}{2}-\frac{J_{\rm y}}{4} & -\frac{\delta_{\rm y}}{2\sqrt{2}} \\
  -\frac{\delta_{\rm y}}{2\sqrt{2}} & -\frac{\delta_{\rm y}}{2\sqrt{2}} & -\frac{J_{\rm y}}{2} 
  \end{array} } \right).
\label{H_triplet_0}
\end{equation}

{We focus on the situation where $\delta_{\rm x,y} \ll  J_{\rm x, y}$. First, we notice that in this limit $\frac{1}{\sqrt{2}}( \ket{T_{12}^{0}T_{34}^{-}} - \ket{T_{12}^{-}T_{34}^{0}})$ is decoupled from the other states. Thus, when the system is diabatically initialized to $\ket{S_{12}T_{34}^{-}}$, it evolves to $\ket{T_{12}^{-}S_{34}}$ and back to $\ket{S_{12}T_{34}^{-}}$ at a frequency $f_{ST}$. To calculate $f_{ST}$, we perform a basis change to}  $\{\ket{0}, \ket{1}, \ket{2} \}  \equiv \{ \frac{1}{\sqrt{2}}(\ket{S_{12} T_{34}^{-}} - \ket{T_{12}^{-} S_{34}}),  \frac{1}{\sqrt{2}}(\ket{S_{12} T_{34}^{-}} + \ket{T_{12}^{-} S_{34}}),  \frac{1}{\sqrt{2}} ( \ket{T_{12}^{0}T_{34}^{-}} - \ket{T_{12}^{-}T_{34}^{0}}) \}$ and separate the Hamiltonian into two terms:

\begin{equation}
H_{T}^{\prime} =
  \left( {\begin{array}{ccc}
    -\frac{J_{\rm x}}{2} & -\frac{\delta_{\rm x}}{2}  & 0 \\
  -\frac{\delta_{\rm x}}{2} & -\frac{J_{\rm x}+J_{\rm y}}{2} & -\frac{\delta_{\rm y}}{2} \\
  0 & -\frac{\delta_{\rm y}}{2} & -\frac{J_{\rm y}}{2} 
  \end{array} } \right) = 
  \left( {\begin{array}{ccc}
    -\frac{J_{\rm x}}{2} & 0  & 0 \\
  0 & -\frac{J_{\rm x}+J_{\rm y}}{2} & 0 \\
  0 & 0 & -\frac{J_{\rm y}}{2} 
  \end{array} } \right) +
    \left( {\begin{array}{ccc}
    0 & -\frac{\delta_{\rm x}}{2}  & 0 \\
  -\frac{\delta_{\rm x}}{2} & 0 & -\frac{\delta_{\rm y}}{2} \\
  0 & -\frac{\delta_{\rm y}}{2} & 0 
  \end{array} } \right)=
  H_{0} + V ,
\label{H_triplet_transformed}
\end{equation} 
 
where $H_0$ only contains diagonal elements $J_{\rm x, y}$ and $V$ only contains off-diagonal elements $\delta_{\rm x,y}$. In the non-degenerate case $\delta_{\rm x,y} \lesssim |J_{\rm x} - J_{\rm y}|$, we apply the second order perturbation theory to the term $V$. The eigenenergies become: 
 
\begin{equation}
E_0 = 
E_{0}^{(0)} + E_{0}^{(1)} + E_{0}^{(2)} = 
\bra{0} H_0 \ket{0} + \bra{0} V \ket{0} + 
\sum_{i=\{1,2\}} \frac{ |\bra{i} V \ket{0}|^2 }{E_{0}^{(0)}-E_{i}} = 
-\frac{J_{\rm x}}{2} + \frac{\delta_{\rm x}^2}{2 J_{\rm y}},
\label{E_triplet_0}
\end{equation}  
 
\begin{equation}
E_1 = 
E_{1}^{(0)} + E_{1}^{(1)} + E_{1}^{(2)} = 
\bra{1} H_0 \ket{1} + \bra{1} V \ket{1} + \sum_{i=\{0,2\}} \frac{ |\bra{i} V \ket{1}|^2 }{E_{1}^{(0)}-E_{i}} = 
-\frac{J_{\rm x}+J_{\rm y}}{2} -\frac{\delta_{\rm x}^2}{2 J_{\rm y}} - \frac{\delta_{\rm y}^2}{2 J_{\rm x}},
\label{E_triplet_1}
\end{equation}  

\begin{equation}
E_2 = 
E_{2}^{(0)} + E_{2}^{(1)} + E_{2}^{(2)} = 
\bra{2} H_0 \ket{2} + \bra{2} V \ket{2} + \sum_{i=\{0,1\}} \frac{ |\bra{i} V \ket{2}|^2 }{E_{2}^{(0)}-E_{i}} = 
-\frac{J_{\rm y}}{2} + \frac{\delta_{\rm y}^2}{2 J_{\rm x}}.
\label{E_triplet_2}
\end{equation}

$\ket{S_{12}T_{34}^{-}}=\frac{1}{\sqrt{2}}(\ket{0}+\ket{1})$ and $\ket{T_{12}^{-}S_{34}}=\frac{1}{\sqrt{2}}(\ket{1}-\ket{0})$ . Thus, we infer that $f_{ST}$ corresponds to the energy difference:
\begin{equation}
f_{ST}= E_{0} - E_{1} =  \frac{J_{\rm y}}{2} + \frac{\delta_{\rm x}^2}{ J_{\rm y}} + \frac{\delta_{\rm y}^2}{2 J_{\rm x}}.
\label{E_triplet_01}
\end{equation} 
 
Equation~(\ref{E_triplet_01}) shows that the $S$-$T^-$ oscillation frequency minimum allows to extract the exchange value $J_{\rm y}$.

According to these calculations, when the barrier gate voltages are varied by $\delta V_{\rm x}$ and $ \delta V_{\rm y}$ at fixed evolution time $t$, the constant  ST probability lines should draw  ellipses centered at the voltages where $J_{12}=J_{34}$ and $J_{14}=J_{23}$. One can use this property to equalize the exchange couplings.
\linebreak

The Hamiltonian $H'_{T}$ can also be diagonalized in the degenerate case $J_{\rm x}=J_{\rm y} =J $. The eigenenergies read as: 
\begin{equation}
\begin{cases}
E_{g} = \frac{-3J - \sqrt{J^2+4\delta_{\rm x}^2 + 4 \delta_{\rm y}^2}}{4} 
\approx-J - \frac{ \delta_{\rm x}^2 + \delta_{\rm y}^2}{2J} \\
E_{e_1} = -\frac{J}{2} \\
E_{e_2} = \frac{-3J + \sqrt{J^2+4\delta_{\rm x}^2 + 4 \delta_{\rm y}^2}}{4} 
\approx -\frac{J}{2} +\frac{ \delta_{\rm x}^2 + \delta_{\rm y}^2}{2J}
\end{cases}
\label{Es_triplet_deg}
\end{equation}

\begin{equation}
\begin{cases}
\ket{g} = \frac{
2 \delta_{\rm x}   \ket{0} 
+ (J+\sqrt{ J^2+ 4\delta_{\rm x}^2 + 4\delta_{\rm y}^2  })   \ket{1}
+ 2 \delta_{\rm y} \ket{2}
}{N_{\rm g}}
\approx \ket{1} \\

\ket{e_{1}} = \frac{ 
-\delta_{\rm y}   \ket{0}  + \delta_{\rm x}   \ket{2} 
}{N_{\rm e_1}}
\approx
    \frac{\delta_{\rm y}}{  \sqrt{\delta_{\rm x}^2 + \delta_{\rm y}^2} } \ket{0} + 
\frac{\delta_{\rm x}}{  \sqrt{\delta_{\rm x}^2 + \delta_{\rm y}^2}  }  \ket{2} \\

\ket{e_{2}} = \frac{
2 \delta_{\rm x}   \ket{0} 
+ (J-\sqrt{ J^2+ 4\delta_{\rm x}^2 + 4\delta_{\rm y}^2  })   \ket{1}
+ 2 \delta_{\rm y} \ket{2}
}{N_{\rm e_2}}
\approx
    -\frac{\delta_{\rm x}}{  \sqrt{\delta_{\rm x}^2 + \delta_{\rm y}^2} } \ket{0} + 
\frac{\delta_{\rm y}}{  \sqrt{\delta_{\rm x}^2 + \delta_{\rm y}^2} } \ket{2}
\end{cases}
\label{States_triplet_deg}
\end{equation}

The initialized $\ket{S_{12}T_{34}^-}$ state can be decomposed as:
\begin{equation}
\ket{S_{12}T_{34}^-} = \frac{1}{\sqrt{2}} \ket{g} 
+ \frac{1}{\sqrt{2}} (  \frac{\delta_{\rm y}}{  \sqrt{\delta_{\rm x}^2 + \delta_{\rm y}^2} }\ket{e_1}  
- \frac{\delta_{\rm x}}{  \sqrt{\delta_{\rm x}^2 + \delta_{\rm y}^2} }  \ket{e_2}).
\label{Init_state_triplet_deg}
\end{equation}

As the system evolves, the measured singlet-triplet probability is:
 \begin{equation}
 \begin{aligned}
 P_{ST}(t) = 
|\bra{S_{12}T_{34}^-} 
e^{-i H_{T}^\prime t/\hbar}
\ket{S_{12}T_{34}^-}|^2 = 
|
\frac{1}{2} e^{-i \omega_{\rm g} t} + \frac{1}{2} \frac{ \delta_{\rm y}^2  }{ \delta_{\rm x}^2 + \delta_{\rm y}^2 } e^{-i \omega_{\rm e_1} t} + \frac{1}{2} \frac{ \delta_{\rm x}^2  }{ \delta_{\rm x}^2 + \delta_{\rm y}^2 } e^{-i \omega_{\rm e_2} t} |^2 \\ 
= 
\frac{\delta_{\rm x}^4 + \delta_{\rm x}^2\delta_{\rm y}^2 + \delta_{\rm y}^4}{2 (\delta_{\rm x}^2+\delta_{\rm y}^2)^2} 
+ \frac{\delta_{\rm x}^2}{2(\delta_{\rm x}^2+\delta_{\rm y}^2)} \cos{(\omega_{\rm e_2}-\omega_{\rm g})t} 
+ \frac{\delta_{\rm y}^2}{2(\delta_{\rm x}^2+\delta_{\rm y}^2)} \cos{(\omega_{\rm e_1}-\omega_{\rm g})t} 
+ \frac{\delta_{\rm x}^2 \delta_{\rm y}^2}{2(\delta_{\rm x}^2+\delta_{\rm y}^2)^2} \cos{(\omega_{\rm e_2}-\omega_{\rm e_1})t} \\
=
\frac{\delta_{\rm x}^4 + \delta_{\rm x}^2\delta_{\rm y}^2 + \delta_{\rm y}^4}{2 (\delta_{\rm x}^2+\delta_{\rm y}^2)^2} 
+ \frac{1}{2} \cos{(\frac{\omega_{\rm e_1}+\omega_{\rm e_2}-2\omega_{\rm g}}{2}t)} 
\cos{(\frac{\omega_{\rm e_1}-\omega_{\rm e_2}}{2}t)}
+ \frac{\delta_{\rm x}^2 -\delta_{\rm y}^2}{2(\delta_{\rm x}^2+\delta_{\rm y}^2)} \sin{(\frac{\omega_{\rm e_1}+\omega_{\rm e_2}-2\omega_{\rm g}}{2}t)} 
\sin{(\frac{\omega_{\rm e_1}-\omega_{\rm e_2}}{2}t)} \\
+ \frac{\delta_{\rm x}^2 \delta_{\rm y}^2}{2(\delta_{\rm x}^2+\delta_{\rm y}^2)^2} \cos{(\omega_{\rm e_2}-\omega_{\rm e_1})t}.
\end{aligned}
 \label{State_evolve_triplet_deg}
\end{equation} 

According to equation~(\ref{State_evolve_triplet_deg}), in two special cases (1) when $\delta_{\rm x }=0$, $f_{ST}$ equals to $\frac{J}{2}+\frac{\delta_{\rm y}^2}{2J}$ and (2) when $\delta_{\rm y }=0$, $f_{ST}$ equals to $\frac{J}{2}+\frac{\delta_{\rm x}^2}{J}$. In the general case $\delta_{\rm x,y } \neq 0$, $P_{ST}$ oscillates with three frequencies where two of them are close to each other resulting in a beating. More specifically, $P_{ST}$ oscillates with a fast frequency $\frac{\omega_{\rm e_1}+\omega_{\rm e_2}-2\omega_{\rm g}}{2}=\frac{J}{2}+\frac{3}{4}\frac{\delta_{\rm x}^2}{J}$ while the amplitude is modulated at a lower frequency $\frac{\omega_{\rm e_1}-\omega_{\rm e_2}}{2}=\frac{\delta_{\rm x}^2+\delta_{\rm y}^2}{4J}$. Therefore, as long as $\delta_{\rm x,y}$ remains sufficiently small, such that $\frac{\delta_{\rm x}^2+\delta_{\rm y}^2}{2J}t_ \lesssim \pi$, a frequency minimum still  appears when $\delta_{\rm x,y}=0$ which allows us to extract $J_{\rm x,y}$.

\subsection{Comparison with experiments}

\begin{figure}[h!]
	\centering
	\includegraphics[width=\textwidth]{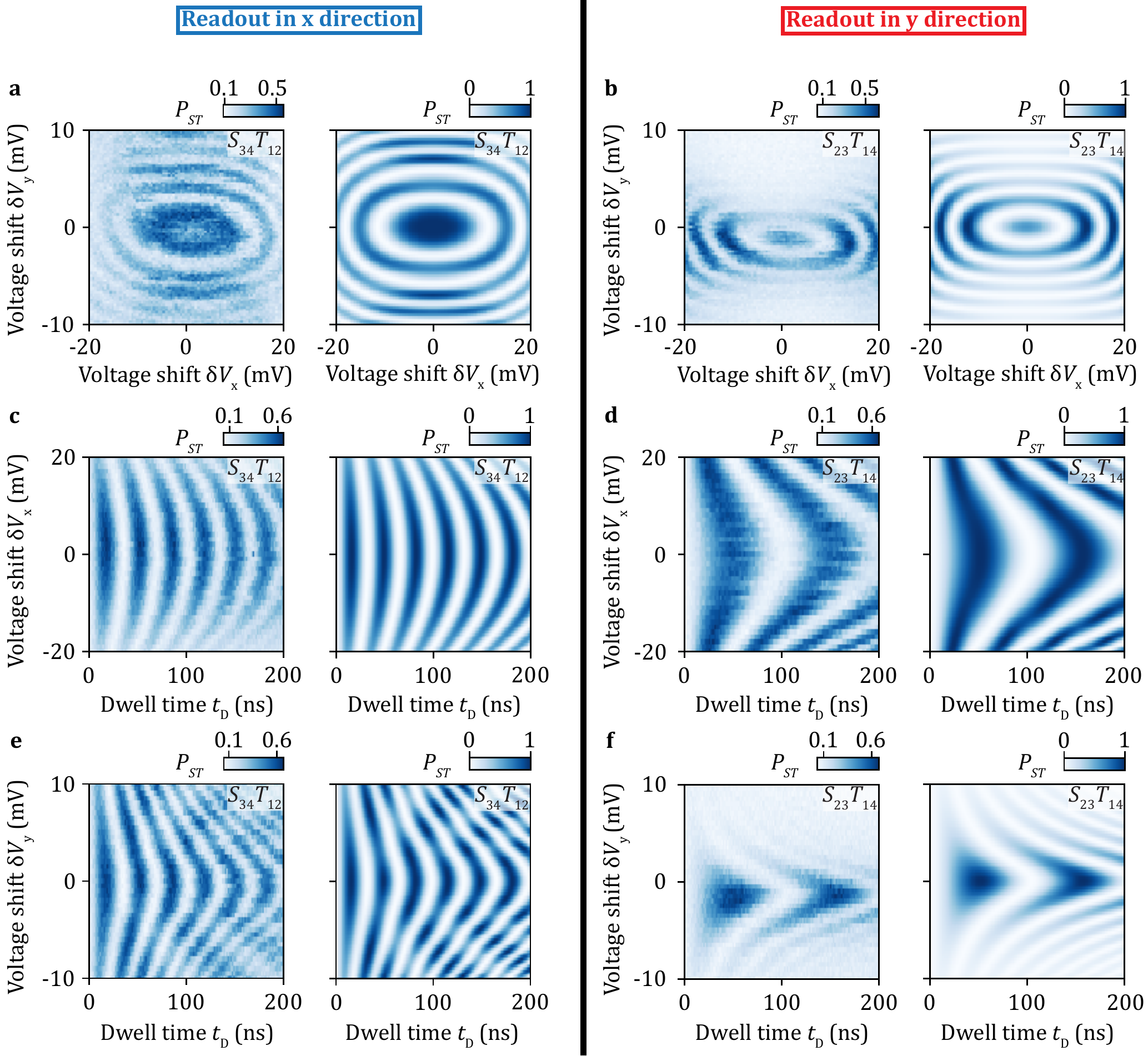}
		\caption{\textbf{Additional measurement data and numerical simulations of four-spin coherent singlet-triplet oscillations at low $\bm{J_{\rm x}}$.} \textbf{a}, \textbf{b}, Probabilities $P_{S_{34}T_{12}}$ and $P_{S_{23}T_{14}}$ as functions of barrier gate voltage variations $\delta V_{\rm x, y}$ at fixed evolution time $t_{\text{D}}=180$~ns.  \textbf{c}, \textbf{d}, Oscillations in $P_{S_{34}T_{12}}$ and $P_{S_{23}T_{14}}$ as functions of gate voltage variation $\delta V_{\rm x}$. \textbf{e}, \textbf{f}, Oscillations in $P_{S_{34}T_{12}}$ and $P_{S_{23}T_{14}}$  as functions of virtual gate voltage variation $\delta V_{\rm y}$. The right panels are numerical simulations based on the Hamiltonian $H_T$. The virtual barrier voltages are varied around the operation point $\rm \{36,-14.1, 20, 5.9\}$~mV which corresponds to the point $\delta V^{\prime}_{\rm x}=20~$mV in Fig.~3.e.  }
	\label{fig:FigSup_ellipse0}
\end{figure}

\begin{figure}[h!]
	\centering
	\includegraphics[width=\textwidth]{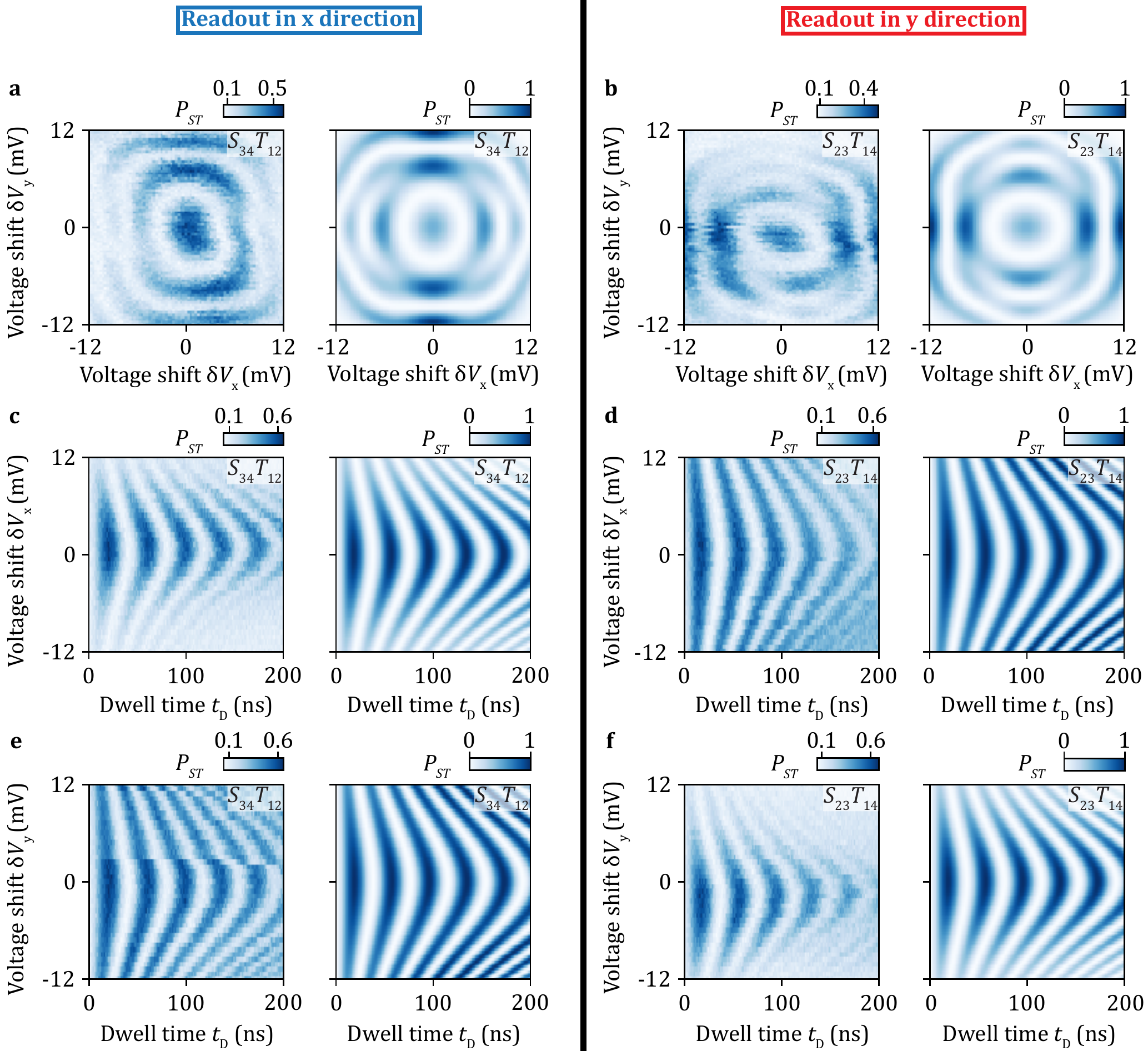}
		\caption{\textbf{Additional measurement data and numerical simulations of four-spin coherent singlet-triplet oscillations at medium $\bm{J_{\rm x}}$.} \textbf{a}, \textbf{b}, Probabilities $P_{S_{34}T_{12}}$ and $P_{S_{23}T_{14}}$ as functions of virtual barrier voltage variations $\delta V_{\rm x, y}$ with a fixed evolution time $t_{\text{D}}=105$~ns. \textbf{c}, \textbf{d}, Oscillations in $P_{S_{34}T_{12}}$ and $P_{S_{23}T_{14}}$ as functions of gate voltage variation $\delta V_{\rm x}$.  \textbf{e}, \textbf{f}, Oscillations in $P_{S_{34}T_{12}}$ and $P_{S_{23}T_{14}}$ as functions of gate voltage variations $\delta V_{\rm y}$.  The right panels are numerical simulations based on the Hamiltonian $H_{T}$. The virtual barrier voltages are varied around $\rm vB_0=\{16,-10.5, 0, 9.5\}~$mV which corresponds to the point $\delta V^{\prime}_{\rm x}=0$~mV in Fig.~3.e.   }
	\label{fig:FigSup_ellipse1}
\end{figure}

\begin{figure}[h!]
	\centering
	\includegraphics[width=\textwidth]{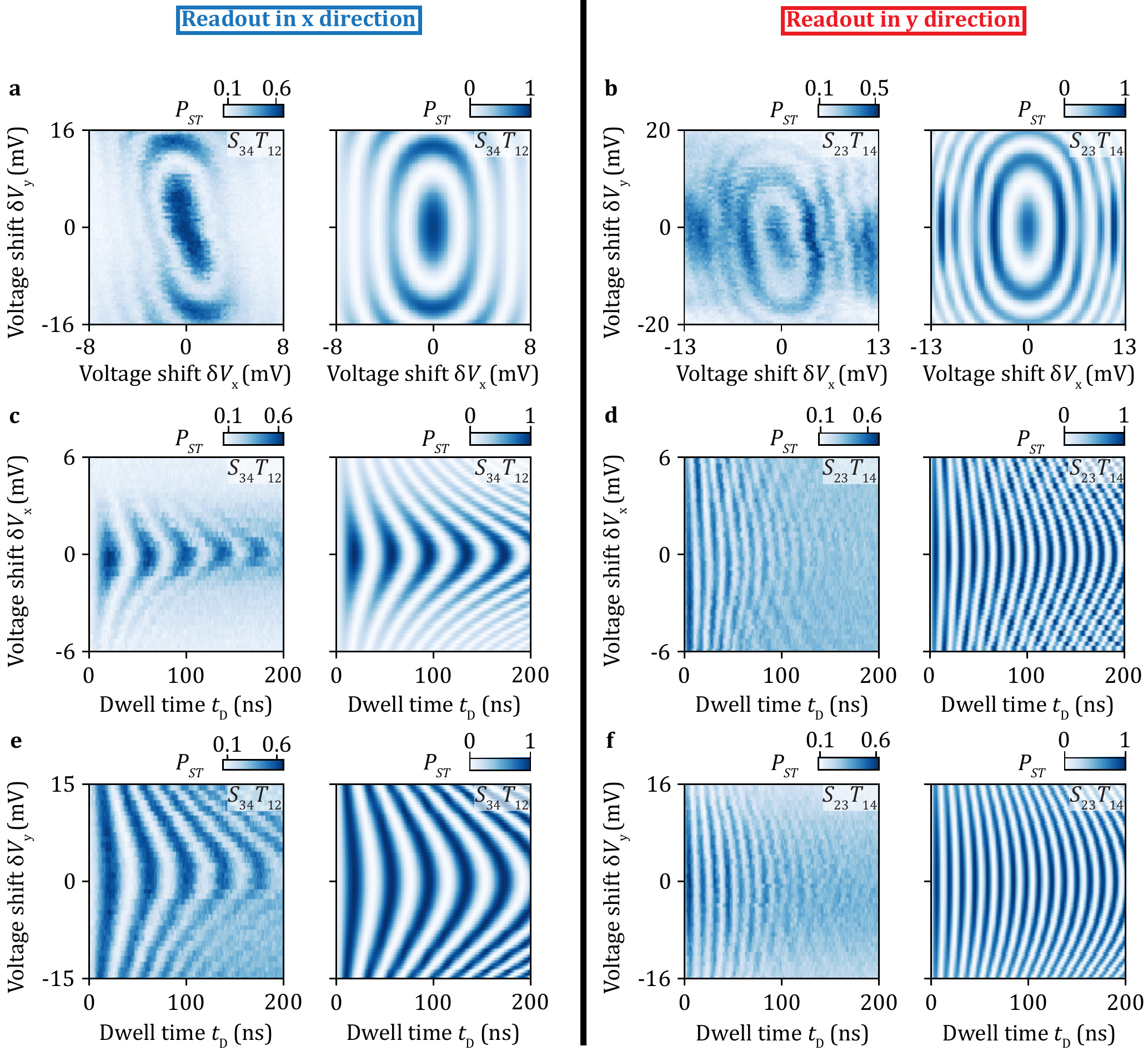}
		\caption{\textbf{Additional measurement data and numerical simulations of four-spin coherent singlet-triplet oscillations at large $\bm{ J_{\rm x}}$.} \textbf{a}, \textbf{b}, Probabilities $P_{S_{34}T_{12}}$ and $P_{S_{23}T_{14}}$ as functions of virtual barrier voltage variations $\delta V_{\rm x, y}$ with a fixed evolution time $t_{\text{D}}=60$~ns. \textbf{c}, \textbf{d} Oscillations in $P_{S_{34}T_{12}}$ and $P_{S_{23}T_{14}}$ probabilities as functions of gate voltage variation $\delta V_{\rm x}$.  \textbf{e, f}, Oscillations in $P_{S_{34}T_{12}}$ and $P_{S_{23}T_{14}}$ as functions of gate voltage variation $\delta V_{\rm y}$.  The panels on the right are numerical simulation based on the Hamiltonian $H_{T}$. The virtual barrier voltages are varied around  $\rm\{-4,-6.9, -20, 13.1\}$~mV, which corresponds to the point $\delta V^{\prime}_{\rm x}=-20$~mV (outside the range in Fig.~3.e).}
	\label{fig:FigSup_ellipse2}
\end{figure}

We perform experiment where we study how global singlet-triplet oscillations evolve when $\delta V_{\rm x}$, $ \delta V_{\rm y}$ and the dwell time $t_{\rm D}$ are varied. Fig.~\ref{fig:FigSup_ellipse0}-\ref{fig:FigSup_ellipse2} show the results of these experiments for different operation points that we compare with numerical simulations of time evolution using the Hamiltonian $H_{T}$.

To perform the simulations, the exchange couplings away from $\delta  V_{\rm x,y} = 0$ are modelled using  exponential models $J_{34/12} = \frac{J_{\rm x} }{2} \exp(\pm\kappa \delta V_{\rm x})$ and $J_{14/23} = \frac{J_{\rm y} }{2} \exp(\pm\kappa \delta V_{\rm y})$ ~\cite{Qiao2020}. The factor $\kappa=0.059$~mV$^{-1}$ is extracted from the frequency of isolated two-spin $S$-$T^-$ oscillations whereas the $J_{\rm x,y}$ values are taken from the frequency minimum in the corresponding sub-figures c-f. The exchange values here are within $10~\%$ of deviation compared to the exchanges displayed in Fig.~3.e.

Fig.~\ref{fig:FigSup_ellipse0}, Fig.~\ref{fig:FigSup_ellipse1}, and  Fig.~\ref{fig:FigSup_ellipse2} show three sets of data/simulations corresponding to experiments where the barrier gate voltages are varied by $\delta  V_{\rm x,y}$ around the points $\{\text{vB}_{12}^0+\delta V^{\prime}_{\text{x}},\text{vB}_{23}^0-\beta\delta V^{\prime}_{\text{x}}, \text{vB}_{34}^0+\delta V^{\prime}_{\text{x}},\text{vB}_{14}^0-\beta\delta V^{\prime}_{\text{x}}\}$ with $\delta V^{\prime}_{\rm x}=$20, 0, -20~mV and $\text{vB}_0=\{\rm vB^0_{12}, vB^0_{34}, vB^0_{23}, vB^0_{14}\}=\{16,-10.5, 0, 9.5\}~$mV ($\text{vB}_0$ is the predetermined set of voltages where exchange couplings are approximately equals mentioned in the main text). These $\delta V^{\prime}_{\rm x}$ values correspond approximately to the center and the limits of the range spanned in Fig.~3.e. In the three cases, the data and the simulations show an overall remarkably good agreement.

Fig.~\ref{fig:FigSup_ellipse0}.a-b, Fig.~\ref{fig:FigSup_ellipse1}.a-b and Fig.~\ref{fig:FigSup_ellipse2}.a-b shows the results of these measurements consisting in varying $\delta V_{\rm x}$, $ \delta V_{\rm y}$ at fixed $t_{\rm D}$ and  the corresponding simulations. We observe that constant probability lines form a network of stripes drawing ellipses centered around $\delta  V_{\rm x,y}\simeq 0~$mV in agreement with the above discussion.

 Fig.~\ref{fig:FigSup_ellipse0}.c-f, Fig.~\ref{fig:FigSup_ellipse1}.c-f  and Fig.~\ref{fig:FigSup_ellipse2}.c-f display the time evolution of global singlet-triplet probabilities measured while varying either $\delta V_{\rm x}$ or $\delta V_{\rm y}$ around the centers of these ellipses. It confirms that the frequency minimum is reached when $\delta  V_{\rm x,y}\simeq 0~$mV i.e. at the center of the ellipse.
 
\begin{figure}[h!]
	\centering
	\includegraphics[width=\textwidth]{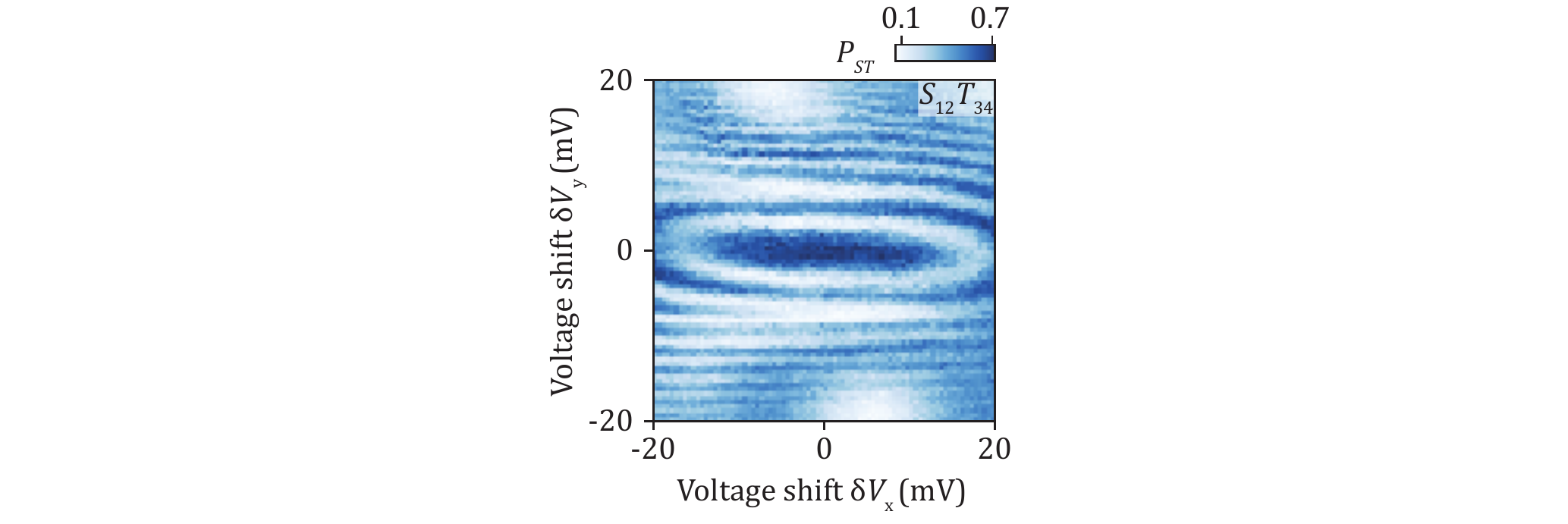}
		\caption{\textbf{Signature of equal exchange couplings.} Probability $P_{S_{12}T_{34}}$ as a function of the voltage variation $\delta  V_{\rm x,y}$ at fixed $t_{\rm D}=113~$ns. The barrier gate voltages are varied around $\{42,-15.18, 26, 4.82\}$~mV which corresponds to the extreme point $\delta V^{\prime}_{\rm x}=26~$mV in Fig.~3.e. The ellipse drawn by constant probability lines is centered around $\delta  V_{\rm x,y}\simeq 0$ indicating that at this point $J_{12}\simeq J_{34}$ and $J_{23}\simeq J_{14}$.}
	\label{fig:FigSup_ellipse26mV}
\end{figure}

In these three configurations, we clearly have $J_{12}\simeq J_{34}$ and $J_{14}\simeq J_{23}$ for $\delta  V_{\rm x,y}\simeq 0~$mV. Similarly for  $\delta V^{\prime}_{\rm x}=26$~mV, singlet-triplet probabilities also draw an ellipse centered on $\delta  V_{\rm x,y}\simeq 0~$mV (Fig.~\ref{fig:FigSup_ellipse26mV}). Extrapolating these results, we assume that these equalities remain valid over the full range of voltage  $-20~\text{mV}\leqslant\delta V^{\prime}_{\rm x}\leqslant 26$~mV spanned in Fig.~3.e and thus that the frequency of global ST oscillations $f_{ST}$ directly gives $J_{\rm x,y}/2$.

We note that the ellipses are tilted and even distorted especially when $J_{\rm x}$ is large. It could indicate a cross-talk between vertical and horizontal virtual barrier gates but also a deviation from the equations derived above which are valid only for small variations of exchange couplings.

In Fig.~\ref{fig:FigSup_ellipse1}.b (initialization/readout in y direction), we also remark that both the measurements and the simulation show a complex pattern when voltages are varied away from the origin. This pattern appears less clearly in the measurement data of Fig.~\ref{fig:FigSup_ellipse1}.a (initialization/readout in x direction).

Likewise, we notice beating patterns in Fig.~\ref{fig:FigSup_ellipse0}.e and  Fig.~\ref{fig:FigSup_ellipse2}.d. They result from the third level, $\frac{1}{\sqrt{2}} (\ket{T_{12}^{0}T_{34}^{-}} - \ket{T_{12}^{-}T_{34}^{0}})$, that is not completely decoupled and has an overlap with the initial state.

\newpage

\newpage

\newpage
\clearpage

\subsection{Uncertainty on the exchange coupling values}

Our method to determine the barrier gate voltages required to have $J_{12}=J_{34}$ and $J_{23}=J_{14}$ leads to some uncertainties on the values of the exchange couplings. They originate from the uncertainty on the determination of the center of ellipse drawn by the oscillations when varying $\delta  V_{\rm x,y}$ at fixed $t$ (see Fig.~\ref{fig:FigSup_ellipse26mV}-\ref{fig:FigSup_ellipse2}). We estimate that the center's position can be determined with a $\pm 2$~mV precision. Small drifts between experiments also typically leads to such uncertainties. Using the exponential models described in the previous subsection, we can then translate this uncertainty into an uncertainty on the exchange values at a given set of barrier gate voltages.

To express it, we assume that at $\delta  V_{\rm x,y} = 0$, the parallel exchange couplings are actually imbalance ($J_{12}\neq J_{34}$ and $J_{14}\neq J_{23}$). We note  $\delta V_{\rm x,y}^0$ the voltage shifts required to reach the balance $J_{12}=J_{34}=J_{\rm x}^0/2$ and $J_{14}=J_{23}=J_{\rm y}^0/2$. Then, as $\kappa\vert\delta  V_{\rm x,y}-\delta V_{\rm x,y}^0 \vert \lesssim 0.1$, we can write  the differences of exchange couplings as:
\begin{equation}
\begin{cases}
\delta_{\rm x } = J_{12}-J_{34} 
= \frac{J^0_{\rm x} }{2} \exp(-\kappa (\delta V_{\rm x}-\delta V_{\rm x}^0)) - \frac{J^0_{\rm x} }{2} \exp(\kappa(\delta V_{\rm x}-\delta V_{\rm x}^0)) 
\simeq J^0_{\rm x} \kappa (\delta V_{\rm x}^0-\delta V_{\rm x}) 
\\
\delta_{\rm y } = J_{23}-J_{14} 
= \frac{J^0_{\rm y} }{2} \exp(-\kappa(\delta V_{\rm y}-\delta V_{\rm y}^0)) - \frac{J^0_{\rm y} }{2} \exp(\kappa (\delta V_{\rm y}-\delta V_{\rm y}^0)) 
\simeq J^0_{\rm y} \kappa (\delta V_{\rm y}^0-\delta V_{\rm y}) 
\end{cases}
.
\label{delta_xy_error_imbalance}
\end{equation}

Likewise, the sums of the exchange couplings are given by:
\begin{equation}
\begin{cases}
J_{\rm x } = J_{12}+J_{34} \simeq J^0_{\rm x}(1+ \frac{\kappa^2(\delta V_{\rm x}-\delta V_{\rm x}^0)^2}{2})
\\
J_{\rm y } = J_{14}+J_{24} \simeq J^0_{\rm y}(1+ \frac{\kappa^2 (\delta V_{\rm y}-\delta V_{\rm y}^0)^2}{2})
\end{cases}
.
\label{sum_xy_error_imbalance}
\end{equation} 

Using equation~(\ref{E_triplet_01}), the errors on the exchange couplings extracted are then given by:
\begin{equation}
\begin{cases}
\sigma_{J_{\rm y}}=2f_{ST} - J_{\rm y}=2\frac{(J_{\rm x}^0)^2}{ J_{\rm y}^0} \kappa^2 {(\delta V_{\rm x}-\delta V_{\rm x}^0)}^2 + \frac{(J_{\rm y}^0)^2}{J_{\rm x}^0} \kappa^2 {(\delta V_{\rm y}-\delta V_{\rm y }^0)}^2
\\
\sigma_{J_{\rm x}} 
= 2\frac{(J_{\rm y}^0)^2}{ J_{\rm x}^0} \kappa^2 {(\delta V_{\rm y}-\delta V_{\rm y}^0)}^2 + \frac{(J_{\rm x}^0)^2}{J_{\rm y}^0}\kappa^2 {(\delta V_{\rm x}-\delta V_{\rm x}^0)}^2
\end{cases}
.
\label{Jxy_error_imbalance}
\end{equation}

Notably, the uncertainty on the position of the center of the ellipse  can only lead to an overestimation of the exchange values. The latter is about 3~MHz in average for the data displayed in Fig.~3.f. Additionally, fitting the four spin singlet-triplet oscillations also leads to uncertainty on the value of $f_{ST}$ from which the exchanges are inferred. We thus assume that the precision on the determination of the exchange couplings is set by the maximum of the two above uncertainties and we use it to draw the error bars in Fig.~3.f. Typically the former uncertainty is much larger than the uncertainty on the frequency fit.

The errors bars on the predicted singlet-singlet oscillation frequency $f_{SS}$ (Fig~4.e),  visibilities  $\mathcal{V}_{\rm x,y}$ (Fig~4.f) and on the singlet-singlet probabilities in the RVB ground state (Fig~5.c) are then drawn by computing the minimum and maximum values of these quantities in the exchange coupling ranges fixed by the uncertainties on $J_{\rm x,y}$.

\section{Limits of the theoretical descriptions}

Up to now, we assumed that the system dynamics is only governed by the Heisenberg Hamiltonian. Yet, the effective Hamiltonian $H_{\rm tot}$ of the system contains other terms. When each quantum dot contains one hole, $H_{\rm tot}$ can be written as:\begin{equation}
H_{\rm tot}=H_{\rm J}+H_{\rm Z}+H_{\rm SO}+H_{\rm hp},
\label{Total_H}
\end{equation}
where $H_{\rm J}=\sum_{\rm i\neq j} J_{\rm ij}(\vec{S_{\rm i} } \cdot \vec{S_{\rm j}}-\frac{1}{4})$, $H_{\rm Z}=\sum_{\rm i} g_{\rm i} \mu_{\rm B} B$, $H_{\rm SO}$ and $H_{\rm hp}$ are respectively the Heisenberg, Zeeman, spin-orbit and hyperfine terms.  The physics of RVB states is determined by $H_{\rm J}$. Thus, we operate only in few milli-Teslas in-plane magnetic fields ensuring that the exchange couplings are the largest energy scales. Thus we can assume that $H_{\rm tot}\simeq H_{\rm J}$ and derive analytical formulas describing the system dynamic. To further justify this approximation, we quantify here the magnitudes of the other energy scales. 

\subsection{Zeeman coupling}

If  $H_{\rm Z}$ is no longer neglected then the states with different total spin states become coupled. In particular, the global singlet states become coupled to the unpolarized triplet state because of the $g$-factor differences between the quantum dots. Using $\{ \ket{S_{12}T_{34}^{0}} , \ket{T_{12}^{0}S_{34}}, \frac{1}{\sqrt{2}}( \ket{T_{12}^{+}T_{34}^{-}} - \ket{T_{12}^{-}T_{34}^{+}})\}=\{2_{T^-},1_{T^-},0_{T^-}\}$, the basis of the corresponding triplet subspace and noting $\{ \ket{0_S}, \ket{1_S} \}=\{    \ket{S_{12}S_{34}}, \frac{1}{\sqrt{3}}(\ket{T^+_{12}T^-_{34}}+
\ket{T^-_{12}T^+_{34}}-\ket{T^0_{12}T^0_{34}}) \}$ the basis of singlet subspace, it can be shown that:

\begin{equation}
\braket{0_{T^-}\vert H_{\rm Z} \vert 0_S} = \frac{1}{2}B\mu_{\rm B}(g_3-g_4),  
\end{equation}

\begin{equation}
\braket{1_{T^-}\vert H_{\rm Z} \vert 0_S} = \frac{1}{2}B\mu_{\rm B}(g_1-g_2),  
\end{equation}

\begin{equation}
\braket{2_{T^-}\vert H_{\rm Z}\vert 0_S} =0,   
\end{equation}

\begin{equation}
\braket{0_{T^-}\vert H_{\rm Z}\vert 1_S} = \frac{1}{2\sqrt{3}}B\mu_{\rm B}(g_4-g_3),   
\end{equation}

\begin{equation}
\braket{1_{T^-}\vert H_{\rm Z}\vert 1_S} = \frac{1}{2\sqrt{3}}B\mu_{\rm B}(g_2-g_1)   
\end{equation}

\begin{equation}
\braket{2_{T^-}\vert H_{\rm Z}\vert 1_S}= \frac{1}{\sqrt{6}}B\mu_{\rm B}(g_1+g_2-g_3-g_4).   
\end{equation}

These coupling terms lead to leakage from the global singlet-singlet subspace leading to discrepancies of the singlet-singlet probabilities measured when looking at the valence bond resonances or eigenstate preparation.

Likewise, the Zeeman term also leads to both additional diagonal and non-diagonal terms in the Hamiltonian describing the dynamics in the $m_S=-1$ global triplet subspace. They read as:

\begin{equation}
\braket{0_{T^-}\vert H_{\rm Z} \vert 0_{T^-}} = - \frac{1}{2}B\mu_{\rm B}(g_3+g_4), 
\end{equation}

\begin{equation}
\braket{1_{T^-}\vert H_{\rm Z} \vert 0_{T^-}} = 0,   \end{equation}

\begin{equation}
 \braket{2_{T^-}\vert H_{\rm Z} \vert 0_{T^-}} = \frac{1}{2\sqrt{2}}B\mu_{\rm B}(g_1-g_2),   \end{equation}

\begin{equation}
\braket{1_{T^-}\vert H_{\rm Z} \vert 1_{T^-}} = -\frac{1}{2}\mu_{\rm B}(g_1+g_2),   \end{equation}

\begin{equation}
\braket{2_{T^-}\vert H_{\rm Z} \vert 1_{T^-}} = \frac{1}{2\sqrt{2}}B\mu_{\rm B}(g_4-g_3),   \end{equation}

\begin{equation}
\braket{2_{T^-}\vert H_{\rm Z} \vert 2_{T^-}}=- \frac{1}{4}B\mu_{\rm B}(g_1+g_2+g_3+g_4).
\end{equation}

These terms can impact the precision and reliability of the extractions of the $J_{\rm ij}$ based on the frequency of global singlet-triplet oscillations.

Finally the Zeeman terms also couples states of the $m_S=-1$ global triplet subspace to a state of the quintuplet subspace $\ket{Q^-}=\frac{1}{\sqrt{2}}(\ket{T_{12}^{0}T_{34}^{-}} + \ket{T_{12}^{-}T_{34}^{0}})$. It leads to three additional couplings terms :

\begin{equation}
\braket{Q^-\vert H_{\rm Z} \vert 2_{T^-}}= \frac{1}{4}B\mu_{\rm B}(g_1+g_2-g_3+g_4),
\end{equation}

\begin{equation}
\braket{Q^-\vert H_{\rm Z} \vert 1_{T^-}}= \frac{1}{2\sqrt{2}}B\mu_{\rm B}(g_3-g_4),
\end{equation}

\begin{equation}
\braket{Q^-\vert H_{\rm Z} \vert 0_{T^-}}= \frac{1}{2\sqrt{2}}B\mu_{\rm B}(g_1-g_2).
\end{equation}

From the Zeeman resonance frequencies measured at 0.65~T, we can infer $g$-factor values in the four dot devices $g_1\simeq 0.14$ , $g_2\simeq 0.24$, $g_3\simeq0.23$ and $g_4\simeq 0.26$, consistent with that measured in ref.~\cite{Hendrickx2021}. The Zeeman energy scale at 1~mT is then typically  smaller than 4~MHz, one order of magnitude below that of exchange interactions justifying that $H_{\rm Z}\ll H_{\rm J}$.

\subsection{Spin-orbit coupling}

In a double quantum dot system, the spin flip terms lead to a coupling between the $\ket{T^{\pm}}$ states and the singlet state and thus to additional leakages outside the subspaces of operation. Spin-orbit couplings also affect the spin-coherence of system by coupling spin states to charge states.

\begin{figure}[h!]
	\centering
	\includegraphics[width=\textwidth]{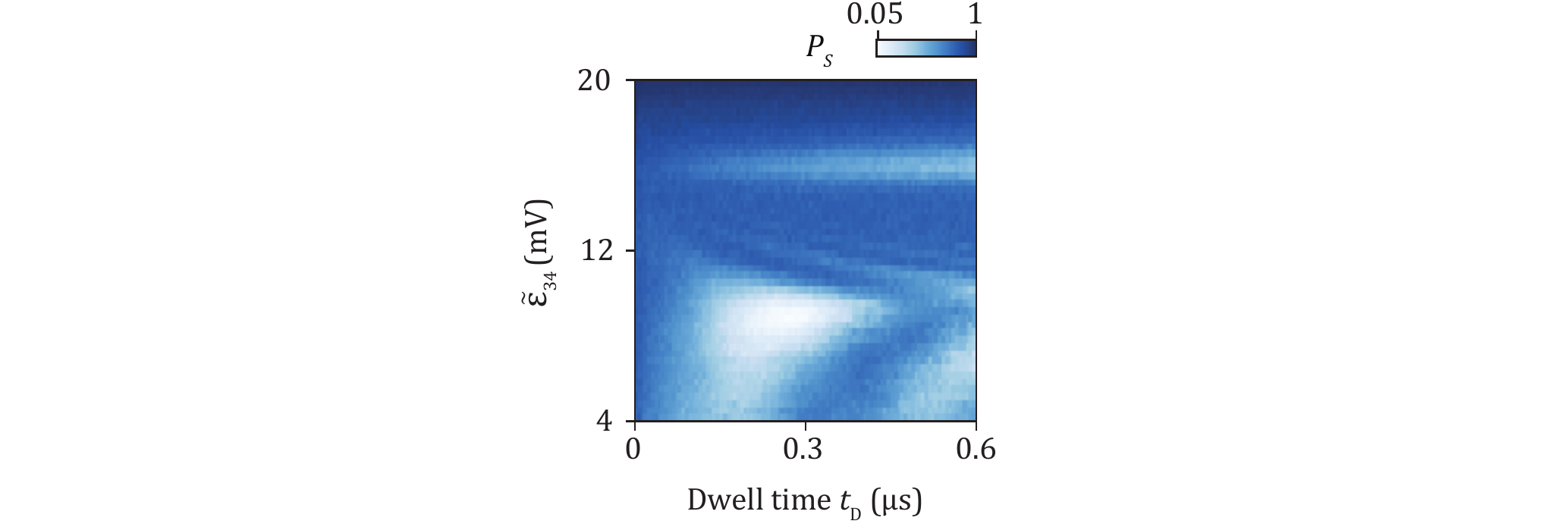}
		\caption{\textbf{$\mathbf{S}$-$\bm{T^-}$ anticrossing at 1 mT}. Singlet-triplet oscillations measured as a function of $\rm \tilde{\varepsilon}_{34}= 0.01\, vP_1 + 0.16\, vP_2 + 1.16\, vP_3 - 0.95\, vP_4$ with Q$_3$Q$_4$ pair at 1~mT. The anticrossing between $S$-$T^-$ states occurs for $\tilde{\varepsilon}_{34}\simeq9$~mV. The frequencies at that point is about 2~MHz.}
	\label{fig:Fig_Sup_Anticrossing_1mT}
\end{figure}

The strength of the spin-flip terms can be evaluated from the minimum of the frequency of $S$-$T^-$ oscillations. Performing similar experiments than that displayed in Fig.~2.b at 1~mT (Fig.~\ref{fig:Fig_Sup_Anticrossing_1mT}), we find that the energy gap at the anticrossing between $S$ and $T^-$ is about 2~MHz. It gives an upper bound to the spin-orbit splitting~\cite{Jirovec2021b}  and confirms that $H_{\rm SO}\ll H_{\rm J}$.

\subsection{Hyperfine fields}

Finally, we focus on the effects of hyperfine interaction. It results in a Zeeman-like term that randomly varies in time, thus to additional couplings between different total spin subspace and additional leakages. Yet, in ref.~\cite{Jirovec2021a}, the authors evaluated the Zeeman energy noise in germanium ST qubits placed in a perpendicular magnetic field. They found $\delta E_{\rm Z}<2~\text{neV}=0.48$~MHz~\cite{Jirovec2021b}, suggesting that the hyperfine interaction can be safely discarded in the theoretical description.

\section{Adiabatic initialization of the s-wave state}

\begin{figure}[H]
	\centering
	\includegraphics[width=\textwidth]{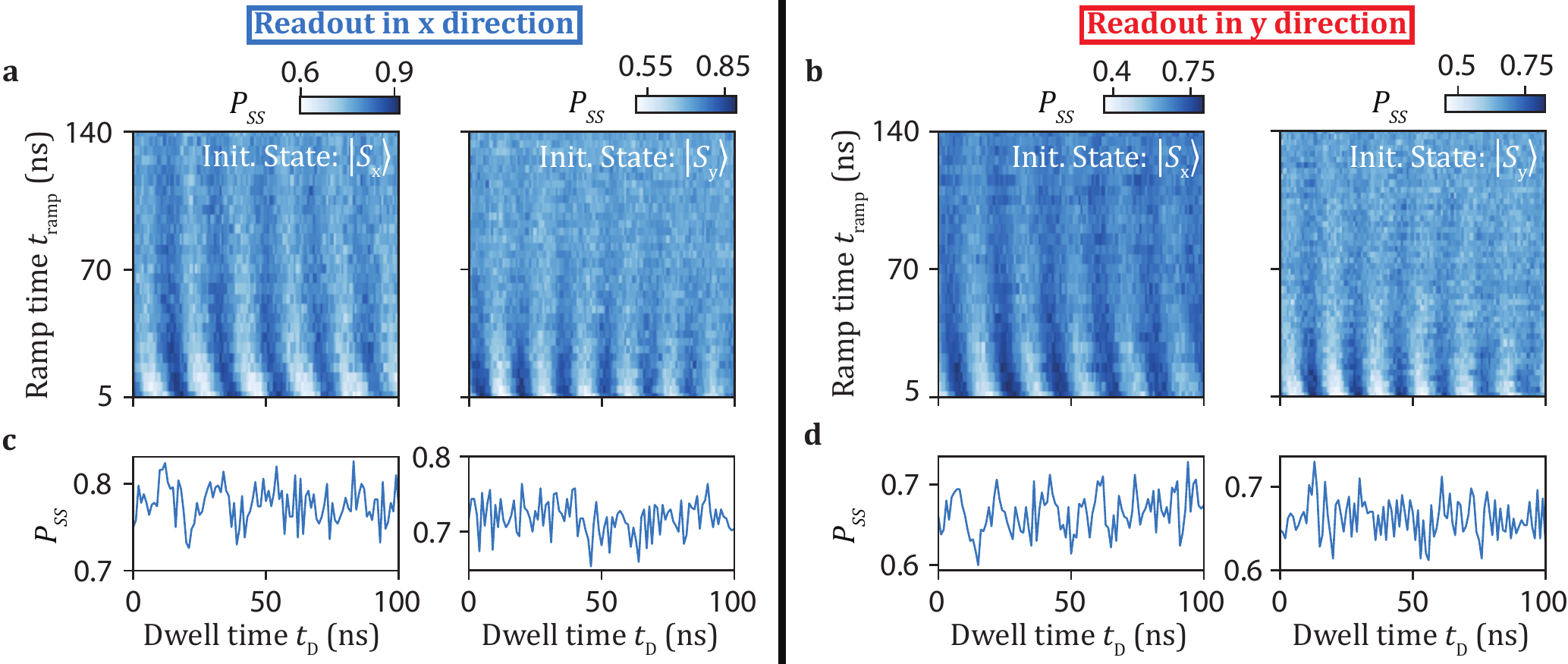}
		\caption{\textbf{Adiabatic preparations of $\mathbf{s}$-wave state.} \textbf{a}, \textbf{b}, Evolution of $P_{\rm S_{12}S_{34}}$, respectively $P_{\rm S_{23}S_{14}}$ oscillations, as functions of $t_{\rm ramp}$ for both initial $\ket{S_{\rm x}}$ and $\ket{S_{\rm y}}$ states. \textbf{c}, \textbf{d}, Linecuts for $t_{\rm ramp}=140$, 141, 140 and 140~ns showing virtual absence of singlet-singlet oscillations.}
	\label{fig:Fig_Sup_Adiabatic_swave}
\end{figure}

Here, we present experiments where a $\ket{s}$ state is prepared by adiabatically equalizing the exchange couplings starting from either a $\ket{S_{\rm x}}$  or $\ket{S_{\rm y}}$ state. The results are shown in Fig.~\ref{fig:Fig_Sup_Adiabatic_swave}.a-b that display the evolution of global singlet-singlet oscillations measured respectively in the x and y direction as functions of the ramp time $t_{\rm ramp}$ used to set the exchange couplings at approximately equal values. In all experiments, we observe a progressive blurring of the oscillations as $t_{\rm ramp}$ is increased resulting from the increased degree of adiabaticity of the pulse. 

For $t_{\rm ramp}\simeq140$~ns, the oscillations virtually disappear as shown by the linecuts of Fig.~\ref{fig:Fig_Sup_Adiabatic_swave}.c-d. It demonstrates that an eigenstate of the system is initialized. For experiments starting from a $\ket{S_{\rm x}}$ state, the average probabilities at $t_{\rm ramp}\simeq140$~ns are $P_{\rm S_{12}S_{34}}=0.78$ and $P_{\rm S_{23}S_{12}}=0.66$. Similarly, for experiments starting from a $\ket{S_{\rm y}}$ state, the average probabilities at $t_{\rm ramp}\simeq140$~ns are $P_{\rm S_{12}S_{34}}=0.72$ and $P_{\rm S_{23}S_{12}}=0.66$. These values are in good agreement with the theoretical expectations for the $s$-wave states: ${\vert\braket{S_{\rm x,y}\vert s}\vert}^2=3/4$.

\bibliography{bib_RVB}